\documentstyle[]{laa} 
\def\PsfigVersion{1.9}
\ifx\undefined\psfig\else\endinput\fi

\let\LaTeXAtSign=\@
\let\@=\relax
\edef\psfigRestoreAt{\catcode`\@=\number\catcode`@\relax}
\catcode`\@=11\relax
\newwrite\@unused
\def\ps@typeout#1{{\let\protect\string\immediate\write\@unused{#1}}}
\ps@typeout{psfig/tex \PsfigVersion}

\def\figurepath{./}

\def\@nnil{\@nil}
\def\@empty{}
\def\@psdonoop#1\@@#2#3{}
\def\@psdo#1:=#2\do#3{\edef\@psdotmp{#2}\ifx\@psdotmp\@empty \else
\expandafter\@psdoloop#2,\@nil,\@nil\@@#1{#3}\fi}
\def\@psdoloop#1,#2,#3\@@#4#5{\def#4{#1}\ifx #4\@nnil \else
#5\def#4{#2}\ifx #4\@nnil \else#5\@ipsdoloop #3\@@#4{#5}\fi\fi}
\def\@ipsdoloop#1,#2\@@#3#4{\def#3{#1}\ifx #3\@nnil
\let\@nextwhile=\@psdonoop \else
#4\relax\let\@nextwhile=\@ipsdoloop\fi\@nextwhile#2\@@#3{#4}}
\def\@tpsdo#1:=#2\do#3{\xdef\@psdotmp{#2}\ifx\@psdotmp\@empty \else
\@tpsdoloop#2\@nil\@nil\@@#1{#3}\fi}
\def\@tpsdoloop#1#2\@@#3#4{\def#3{#1}\ifx #3\@nnil
\let\@nextwhile=\@psdonoop \else
#4\relax\let\@nextwhile=\@tpsdoloop\fi\@nextwhile#2\@@#3{#4}}
\ifx\undefined\fbox
\newdimen\fboxrule
\newdimen\fboxsep
\newdimen\ps@tempdima
\newbox\ps@tempboxa
\long\def\fbox#1{\leavevmode\setbox\ps@tempboxa\hbox{#1}\ps@tempdima\fboxrule
\advance\ps@tempdima \fboxsep \advance\ps@tempdima \dp\ps@tempboxa
\hbox{\lower \ps@tempdima\hbox
{\vbox{\hrule height \fboxrule
\hbox{\vrule width \fboxrule \hskip\fboxsep
\vbox{\vskip\fboxsep \box\ps@tempboxa\vskip\fboxsep}\hskip
\fboxsep\vrule width \fboxrule}
\hrule height \fboxrule}}}}
\fi
\newread\ps@stream
\newif\ifnot@eof       
\newif\if@noisy        
\newif\if@atend        
\newif\if@psfile       
{\catcode`\%=12\global\gdef\epsf@start{
\def\epsf@PS{PS}
\def\epsf@getbb#1{%
\openin\ps@stream=#1
\ifeof\ps@stream\ps@typeout{Error, File #1 not found}\else
{\not@eoftrue \chardef\other=12
\def\do##1{\catcode`##1=\other}\dospecials \catcode`\ =10
\loop
\if@psfile
	  \read\ps@stream to \epsf@fileline
\else{
	  \obeyspaces
\read\ps@stream to \epsf@tmp\global\let\epsf@fileline\epsf@tmp}
\fi
\ifeof\ps@stream\not@eoffalse\else
\if@psfile\else
\expandafter\epsf@test\epsf@fileline:. \\%
\fi
\expandafter\epsf@aux\epsf@fileline:. \\%
\fi
\ifnot@eof\repeat
}\closein\ps@stream\fi}%
\long\def\epsf@test#1#2#3:#4\\{\def\epsf@testit{#1#2}
			\ifx\epsf@testit\epsf@start\else
\ps@typeout{Warning! File does not start with `\epsf@start'.  It may not be a
PostScript file.}
			\fi
			\@psfiletrue} 
{\catcode`\%=12\global\let\epsf@percent=
\long\def\epsf@aux#1#2:#3\\{\ifx#1\epsf@percent
\def\epsf@testit{#2}\ifx\epsf@testit\epsf@bblit
	\@atendfalse
\epsf@atend #3 . \\%
	\if@atend
	   \if@verbose{
		\ps@typeout{psfig: found `(atend)'; continuing search}
	   }\fi
\else
\epsf@grab #3 . . . \\%
\not@eoffalse
\global\no@bbfalse
\fi
\fi\fi}%
\def\epsf@grab #1 #2 #3 #4 #5\\{%
\global\def\epsf@llx{#1}\ifx\epsf@llx\empty
\epsf@grab #2 #3 #4 #5 .\\\else
\global\def\epsf@lly{#2}%
\global\def\epsf@urx{#3}\global\def\epsf@ury{#4}\fi}%
\def\epsf@atendlit{(atend)}
\def\epsf@atend #1 #2 #3\\{%
\def\epsf@tmp{#1}\ifx\epsf@tmp\empty
\epsf@atend #2 #3 .\\\else
\ifx\epsf@tmp\epsf@atendlit\@atendtrue\fi\fi}

\chardef\psletter = 11 
\chardef\other = 12

\newif \ifdebug 
\newif\ifc@mpute 
\c@mputetrue 

\let\then = \relax
\def\r@dian{pt }
\let\r@dians = \r@dian
\let\dimensionless@nit = \r@dian
\let\dimensionless@nits = \dimensionless@nit
\def\internal@nit{sp }
\let\internal@nits = \internal@nit
\newif\ifstillc@nverging
\def \Mess@ge #1{\ifdebug \then \message {#1} \fi}

{ 
	\catcode `\@ = \psletter
	\gdef \nodimen {\expandafter \n@dimen \the \dimen}
	\gdef \term #1 #2 #3%
	       {\edef \t@ {\the #1}
		\edef \t@@ {\expandafter \n@dimen \the #2\r@dian}%
		\t@rm {\t@} {\t@@} {#3}%
	       }
	\gdef \t@rm #1 #2 #3%
	       {{%
		\count 0 = 0
		\dimen 0 = 1 \dimensionless@nit
		\dimen 2 = #2\relax
		\Mess@ge {Calculating term #1 of \nodimen 2}%
		\loop
		\ifnum	\count 0 < #1
		\then	\advance \count 0 by 1
			\Mess@ge {Iteration \the \count 0 \space}%
			\Multiply \dimen 0 by {\dimen 2}%
			\Mess@ge {After multiplication, term = \nodimen 0}%
			\Divide \dimen 0 by {\count 0}%
			\Mess@ge {After division, term = \nodimen 0}%
		\repeat
		\Mess@ge {Final value for term #1 of
				\nodimen 2 \space is \nodimen 0}%
		\xdef \Term {#3 = \nodimen 0 \r@dians}%
		\aftergroup \Term
	       }}
	\catcode `\p = \other
	\catcode `\t = \other
	\gdef \n@dimen #1pt{#1} 
}

\def \Divide #1by #2{\divide #1 by #2} 

\def \Multiply #1by #2
{{
	\count 0 = #1\relax
	\count 2 = #2\relax
	\count 4 = 65536
	\Mess@ge {Before scaling, count 0 = \the \count 0 \space and
			count 2 = \the \count 2}%
	\ifnum	\count 0 > 32767 
	\then	\divide \count 0 by 4
		\divide \count 4 by 4
	\else	\ifnum	\count 0 < -32767
		\then	\divide \count 0 by 4
			\divide \count 4 by 4
		\else
		\fi
	\fi
	\ifnum	\count 2 > 32767 
	\then	\divide \count 2 by 4
		\divide \count 4 by 4
	\else	\ifnum	\count 2 < -32767
		\then	\divide \count 2 by 4
			\divide \count 4 by 4
		\else
		\fi
	\fi
	\multiply \count 0 by \count 2
	\divide \count 0 by \count 4
	\xdef \product {#1 = \the \count 0 \internal@nits}%
	\aftergroup \product
}}

\def\r@duce{\ifdim\dimen0 > 90\r@dian \then   
		\multiply\dimen0 by -1
		\advance\dimen0 by 180\r@dian
		\r@duce
	    \else \ifdim\dimen0 < -90\r@dian \then  
		\advance\dimen0 by 360\r@dian
		\r@duce
		\fi
	    \fi}

\def\Sine#1%
{{%
	\dimen 0 = #1 \r@dian
	\r@duce
	\ifdim\dimen0 = -90\r@dian \then
	   \dimen4 = -1\r@dian
	   \c@mputefalse
	\fi
	\ifdim\dimen0 = 90\r@dian \then
	   \dimen4 = 1\r@dian
	   \c@mputefalse
	\fi
	\ifdim\dimen0 = 0\r@dian \then
	   \dimen4 = 0\r@dian
	   \c@mputefalse
	\fi
	\ifc@mpute \then
		\divide\dimen0 by 180
		\dimen0=3.141592654\dimen0
		\dimen 2 = 3.1415926535897963\r@dian 
		\divide\dimen 2 by 2 
		\Mess@ge {Sin: calculating Sin of \nodimen 0}%
		\count 0 = 1 
		\dimen 2 = 1 \r@dian 
		\dimen 4 = 0 \r@dian 
		\loop
			\ifnum	\dimen 2 = 0 
			\then	\stillc@nvergingfalse
			\else	\stillc@nvergingtrue
			\fi
			\ifstillc@nverging 
			\then	\term {\count 0} {\dimen 0} {\dimen 2}%
				\advance \count 0 by 2
				\count 2 = \count 0
				\divide \count 2 by 2
				\ifodd	\count 2 
				\then	\advance \dimen 4 by \dimen 2
				\else	\advance \dimen 4 by -\dimen 2
				\fi
		\repeat
	\fi
			\xdef \sine {\nodimen 4}%
}}

\def\Cosine#1{\ifx\sine\UnDefined\edef\Savesine{\relax}\else
		             \edef\Savesine{\sine}\fi
	{\dimen0=#1\r@dian\advance\dimen0 by 90\r@dian
	 \Sine{\nodimen 0}
	 \xdef\cosine{\sine}
	 \xdef\sine{\Savesine}}}

\def\psdraft{
	\def\@psdraft{0}
}
\def\psfull{
	\def\@psdraft{100}
}

\psfull

\newif\if@scalefirst
\def\psscalefirst{\@scalefirsttrue}
\def\psrotatefirst{\@scalefirstfalse}
\psrotatefirst

\newif\if@draftbox
\def\psnodraftbox{
	\@draftboxfalse
}
\def\psdraftbox{
	\@draftboxtrue
}
\@draftboxtrue

\newif\if@prologfile
\newif\if@postlogfile
\def\pssilent{
	\@noisyfalse
}
\def\psnoisy{
	\@noisytrue
}
\psnoisy
\newif\if@bbllx
\newif\if@bblly
\newif\if@bburx
\newif\if@bbury
\newif\if@height
\newif\if@width
\newif\if@rheight
\newif\if@rwidth
\newif\if@angle
\newif\if@clip
\newif\if@verbose
\def\@p@@sclip#1{\@cliptrue}

\newif\if@decmpr

\def\@p@@sfigure#1{\def\@p@sfile{null}\def\@p@sbbfile{null}
	        \openin1=#1.bb
		\ifeof1\closein1
	        	\openin1=\figurepath#1.bb
			\ifeof1\closein1
			        \openin1=#1
				\ifeof1\closein1%
				       \openin1=\figurepath#1
					\ifeof1
					   \ps@typeout{Error, File #1 not found}
						\if@bbllx\if@bblly
				   		\if@bburx\if@bbury
			      				\def\@p@sfile{#1}%
			      				\def\@p@sbbfile{#1}%
							\@decmprfalse
				  	   	\fi\fi\fi\fi
					\else\closein1
				    		\def\@p@sfile{\figurepath#1}%
				    		\def\@p@sbbfile{\figurepath#1}%
						\@decmprfalse
	                       		\fi%
			 	\else\closein1%
					\def\@p@sfile{#1}
					\def\@p@sbbfile{#1}
					\@decmprfalse
			 	\fi
			\else
				\def\@p@sfile{\figurepath#1}
				\def\@p@sbbfile{\figurepath#1.bb}
				\@decmprtrue
			\fi
		\else
			\def\@p@sfile{#1}
			\def\@p@sbbfile{#1.bb}
			\@decmprtrue
		\fi}

\def\@p@@sfile#1{\@p@@sfigure{#1}}

\def\@p@@sbbllx#1{
		\@bbllxtrue
		\dimen100=#1
		\edef\@p@sbbllx{\number\dimen100}
}
\def\@p@@sbblly#1{
		\@bbllytrue
		\dimen100=#1
		\edef\@p@sbblly{\number\dimen100}
}
\def\@p@@sbburx#1{
		\@bburxtrue
		\dimen100=#1
		\edef\@p@sbburx{\number\dimen100}
}
\def\@p@@sbbury#1{
		\@bburytrue
		\dimen100=#1
		\edef\@p@sbbury{\number\dimen100}
}
\def\@p@@sheight#1{
		\@heighttrue
		\dimen100=#1
		\edef\@p@sheight{\number\dimen100}
}
\def\@p@@swidth#1{
		\@widthtrue
		\dimen100=#1
		\edef\@p@swidth{\number\dimen100}
}
\def\@p@@srheight#1{
		\@rheighttrue
		\dimen100=#1
		\edef\@p@srheight{\number\dimen100}
}
\def\@p@@srwidth#1{
		\@rwidthtrue
		\dimen100=#1
		\edef\@p@srwidth{\number\dimen100}
}
\def\@p@@sangle#1{
		\@angletrue
		\edef\@p@sangle{#1} 
}
\def\@p@@ssilent#1{
		\@verbosefalse
}
\def\@p@@sprolog#1{\@prologfiletrue\def\@prologfileval{#1}}
\def\@p@@spostlog#1{\@postlogfiletrue\def\@postlogfileval{#1}}
\def\@cs@name#1{\csname #1\endcsname}
\def\@setparms#1=#2,{\@cs@name{@p@@s#1}{#2}}
\def\ps@init@parms{
		\@bbllxfalse \@bbllyfalse
		\@bburxfalse \@bburyfalse
		\@heightfalse \@widthfalse
		\@rheightfalse \@rwidthfalse
		\def\@p@sbbllx{}\def\@p@sbblly{}
		\def\@p@sbburx{}\def\@p@sbbury{}
		\def\@p@sheight{}\def\@p@swidth{}
		\def\@p@srheight{}\def\@p@srwidth{}
		\def\@p@sangle{0}
		\def\@p@sfile{} \def\@p@sbbfile{}
		\def\@p@scost{10}
		\def\@sc{}
		\@prologfilefalse
		\@postlogfilefalse
		\@clipfalse
		\if@noisy
			\@verbosetrue
		\else
			\@verbosefalse
		\fi
}
\def\parse@ps@parms#1{
	 	\@psdo\@psfiga:=#1\do
		   {\expandafter\@setparms\@psfiga,}}
\newif\ifno@bb
\def\bb@missing{
	\if@verbose{
		\ps@typeout{psfig: searching \@p@sbbfile \space  for bounding box}
	}\fi
	\no@bbtrue
	\epsf@getbb{\@p@sbbfile}
\ifno@bb \else \bb@cull\epsf@llx\epsf@lly\epsf@urx\epsf@ury\fi
}
\def\bb@cull#1#2#3#4{
	\dimen100=#1 bp\edef\@p@sbbllx{\number\dimen100}
	\dimen100=#2 bp\edef\@p@sbblly{\number\dimen100}
	\dimen100=#3 bp\edef\@p@sbburx{\number\dimen100}
	\dimen100=#4 bp\edef\@p@sbbury{\number\dimen100}
	\no@bbfalse
}
\newdimen\p@intvaluex
\newdimen\p@intvaluey
\def\rotate@#1#2{{\dimen0=#1 sp\dimen1=#2 sp
		  \global\p@intvaluex=\cosine\dimen0
		  \dimen3=\sine\dimen1
		  \global\advance\p@intvaluex by -\dimen3
		  \global\p@intvaluey=\sine\dimen0
		  \dimen3=\cosine\dimen1
		  \global\advance\p@intvaluey by \dimen3
		  }}
\def\compute@bb{
		\no@bbfalse
		\if@bbllx \else \no@bbtrue \fi
		\if@bblly \else \no@bbtrue \fi
		\if@bburx \else \no@bbtrue \fi
		\if@bbury \else \no@bbtrue \fi
		\ifno@bb \bb@missing \fi
		\ifno@bb \ps@typeout{FATAL ERROR: no bb supplied or found}
			\no-bb-error
		\fi
		\count203=\@p@sbburx
		\count204=\@p@sbbury
		\advance\count203 by -\@p@sbbllx
		\advance\count204 by -\@p@sbblly
		\edef\ps@bbw{\number\count203}
		\edef\ps@bbh{\number\count204}
		\if@angle
			\Sine{\@p@sangle}\Cosine{\@p@sangle}
	        	{\dimen100=\maxdimen\xdef\r@p@sbbllx{\number\dimen100}
					    \xdef\r@p@sbblly{\number\dimen100}
			                    \xdef\r@p@sbburx{-\number\dimen100}
					    \xdef\r@p@sbbury{-\number\dimen100}}
\def\minmaxtest{
			   \ifnum\number\p@intvaluex<\r@p@sbbllx
			      \xdef\r@p@sbbllx{\number\p@intvaluex}\fi
			   \ifnum\number\p@intvaluex>\r@p@sbburx
			      \xdef\r@p@sbburx{\number\p@intvaluex}\fi
			   \ifnum\number\p@intvaluey<\r@p@sbblly
			      \xdef\r@p@sbblly{\number\p@intvaluey}\fi
			   \ifnum\number\p@intvaluey>\r@p@sbbury
			      \xdef\r@p@sbbury{\number\p@intvaluey}\fi
			   }
			\rotate@{\@p@sbbllx}{\@p@sbblly}
			\minmaxtest
			\rotate@{\@p@sbbllx}{\@p@sbbury}
			\minmaxtest
			\rotate@{\@p@sbburx}{\@p@sbblly}
			\minmaxtest
			\rotate@{\@p@sbburx}{\@p@sbbury}
			\minmaxtest
			\edef\@p@sbbllx{\r@p@sbbllx}\edef\@p@sbblly{\r@p@sbblly}
			\edef\@p@sbburx{\r@p@sbburx}\edef\@p@sbbury{\r@p@sbbury}
		\fi
		\count203=\@p@sbburx
		\count204=\@p@sbbury
		\advance\count203 by -\@p@sbbllx
		\advance\count204 by -\@p@sbblly
		\edef\@bbw{\number\count203}
		\edef\@bbh{\number\count204}
}
\def\in@hundreds#1#2#3{\count240=#2 \count241=#3
		     \count100=\count240	
		     \divide\count100 by \count241
		     \count101=\count100
		     \multiply\count101 by \count241
		     \advance\count240 by -\count101
		     \multiply\count240 by 10
		     \count101=\count240	
		     \divide\count101 by \count241
		     \count102=\count101
		     \multiply\count102 by \count241
		     \advance\count240 by -\count102
		     \multiply\count240 by 10
		     \count102=\count240	
		     \divide\count102 by \count241
		     \count200=#1\count205=0
		     \count201=\count200
			\multiply\count201 by \count100
		 	\advance\count205 by \count201
		     \count201=\count200
			\divide\count201 by 10
			\multiply\count201 by \count101
			\advance\count205 by \count201
		     \count201=\count200
			\divide\count201 by 100
			\multiply\count201 by \count102
			\advance\count205 by \count201
		     \edef\@result{\number\count205}
}
\def\compute@wfromh{
		\in@hundreds{\@p@sheight}{\@bbw}{\@bbh}
		\edef\@p@swidth{\@result}
}
\def\compute@hfromw{
	        \in@hundreds{\@p@swidth}{\@bbh}{\@bbw}
		\edef\@p@sheight{\@result}
}
\def\compute@handw{
		\if@height
			\if@width
			\else
				\compute@wfromh
			\fi
		\else
			\if@width
				\compute@hfromw
			\else
				\edef\@p@sheight{\@bbh}
				\edef\@p@swidth{\@bbw}
			\fi
		\fi
}
\def\compute@resv{
		\if@rheight \else \edef\@p@srheight{\@p@sheight} \fi
		\if@rwidth \else \edef\@p@srwidth{\@p@swidth} \fi
}
\def\compute@sizes{
	\compute@bb
	\if@scalefirst\if@angle
	\if@width
	   \in@hundreds{\@p@swidth}{\@bbw}{\ps@bbw}
	   \edef\@p@swidth{\@result}
	\fi
	\if@height
	   \in@hundreds{\@p@sheight}{\@bbh}{\ps@bbh}
	   \edef\@p@sheight{\@result}
	\fi
	\fi\fi
	\compute@handw
	\compute@resv}

\def\psfig#1{\vbox {
	\ps@init@parms
	\parse@ps@parms{#1}
	\compute@sizes
	\ifnum\@p@scost<\@psdraft{
		\special{ps::[begin] 	\@p@swidth \space \@p@sheight \space
				\@p@sbbllx \space \@p@sbblly \space
				\@p@sbburx \space \@p@sbbury \space
				startTexFig \space }
		\if@angle
			\special {ps:: \@p@sangle \space rotate \space}
		\fi
		\if@clip{
			\if@verbose{
				\ps@typeout{(clip)}
			}\fi
			\special{ps:: doclip \space }
		}\fi
		\if@prologfile
		    \special{ps: plotfile \@prologfileval \space } \fi
		\if@decmpr{
			\if@verbose{
				\ps@typeout{psfig: including \@p@sfile.Z \space }
			}\fi
			\special{ps: plotfile "`zcat \@p@sfile.Z" \space }
		}\else{
			\if@verbose{
				\ps@typeout{psfig: including \@p@sfile \space }
			}\fi
			\special{ps: plotfile \@p@sfile \space }
		}\fi
		\if@postlogfile
		    \special{ps: plotfile \@postlogfileval \space } \fi
		\special{ps::[end] endTexFig \space }
		\vbox to \@p@srheight sp{
			\hbox to \@p@srwidth sp{
				\hss
			}
		\vss
		}
	}\else{
		\if@draftbox{
			\hbox{\frame{\vbox to \@p@srheight sp{
			\vss
			\hbox to \@p@srwidth sp{ \hss \@p@sfile \hss }
			\vss
			}}}
		}\else{
			\vbox to \@p@srheight sp{
			\vss
			\hbox to \@p@srwidth sp{\hss}
			\vss
			}
		}\fi

	}\fi
}}
\psfigRestoreAt
\let\@=\LaTeXAtSign
\newcommand{\be}{\begin{equation}}
\newcommand{\ee}{\end{equation}}
\newcommand{\ti}[1]{\mbox{\tiny{#1}}}
\begin{document}
%
\thesaurus{02(08.15.1; 08.06.3; 12.04.3)}
\title{On improved Cepheid distance estimators}
%
\author{S.M. Kanbur\inst{1}
  \and  M.A. Hendry\inst{2}}
\offprints{S.M. Kanbur} 
\institute{Department of Physics and Astronomy,
          University of Glasgow, Glasgow, UK
  \and Astronomy Centre, University of Sussex, Falmer, Brighton, UK}
\date{Submitted December 1994}
\maketitle
\begin{abstract}
The use of Cepheids as distance indicators on Galactic and
extragalactic distance scales is based upon the Cepheid period -
luminosity (PL) and period - luminosity - colour (PLC) relations.
These relations are usually derived in terms of the properties of Cepheids
at mean light -- i.e. averaged over their pulsation cycle.
In this paper, we derive a physical argument for the existence
of PL and PLC relations at maximum light. We examine in detail a
sample of Cepheids in the Large Magellanic Cloud, and
compare the variance of some PL and PLC type
distance indicators based on mean and maximum light.

We show that for the LMC data considered, a PLC relation based on maximum
light
leads to a distance estimator with a dispersion about $10 \%$ smaller
than its counterpart using mean light.
We also show that for
the LMC, a PLC type relation constructed using observations at both
maximum and mean light
has a significantly $( > 50 \%)$ smaller dispersion than a PLC relation using
either maximum or mean light alone.
A comparable $( > 30 \%)$ reduction in the dispersion of the corresponding
distance estimator, however, in this case requires the relation be applied
to a large $( n > 30)$ group of equidistant Cepheids in, e.g., a distant
galaxy. Recent HST observations of IC4182, M81 and M100 already provide
suitable candidate data sets for this relation. The use of
maximum light in constructing PLC type relations for galactic
and extragalactic Cepheids is, therefore, shown to be
an interesting topic for further study. These investigations are under way.
\end{abstract}
%
%
\section{Introduction}

Cepheids are high luminosity radially pulsating variable stars. Their
intrinsic brightness ranges from $-2 > M_V > -6$ and makes them suitable
candidates for distance indicators on Galactic and extragalactic
distance scales. This is based on the Cepheid period - luminosity (PL)
and period - luminosity - colour (PLC) relations.
Examples of such PL and PLC relations for the Magellanic Clouds
are given in Figure 4 of Madore and Freedman (1991) and in
Figure 3 of Caldwell and Coulson (1986; hereafter CC) respectively.
In these relations
the luminosity, measured by the magnitude, and colour are
mean quantities taken over the pulsational cycle.
Motivated by the work of Simon, Kanbur and Mihalas (1993; hereafter SKM),
in this paper we derive a physical argument for the existence of
PL and PLC relations for Cepheids based on their properties at
{\em maximum\/} light. This argument thus provides a physical
justification for the work of Sandage and Tammann (1968), who introduced
a PL relation at maximum light. However, we have extended their work
by the introduction of a colour term and the simultaneous use of maximum and
mean light.

We then examine in detail a sample
of Cepheids in the Large Magellanic Cloud (LMC),
using multicolour photometric data originally
presented in Martin and Warren (1979) and discussed
in Martin, Warren and Feast (1979; hereafter MWF).
These data have (B-V) colours and were used purely as an
illustration since this was the only data available to us at
the time. The principles of our
analysis are applicable in any wavelength range, however, and the
extension of our analysis to other data sets will form
part of our future work.
Following the statistical formalism described in
Hendry and Simmons (1990, 1994; hereafter HS90, HS94),
we obtain `optimal' (in the sense of unbiased and minimum variance)
distance estimators corresponding to several different
PL and PLC--type relations derived from this calibrating sample.
We show that distance estimators based on the properties of
the Cepheid at maximum light can have significantly
smaller variance than those derived for Cepheids at mean light.

The paper is organised as follows: Sect. 2 describes the theoretical
basis for the PL and PLC relations, originally introduced by
Sandage (1958). Sect. 3 explains the physical
reasoning behind our reformulation and extension of
these relations in terms of Cepheid maximum light and colour.
Sect. 4 outlines briefly how one may
use our new relations to derive the corresponding
Cepheid distance indicators. In the appendix we discuss
the statistical model with which we derive these
distance indicators, and describe the significance tests which
we used to compare the variance of distance estimators defined
for Cepheids at mean and maximum light. In Sect. 5
we describe the LMC Cepheid data upon which our
analysis is based, and test the validity of the assumptions
made in the appendix concerning the statistical properties of this sample.
Sect. 6 presents our results and discussions. In Sect. 7 and 8 we report
our conclusions and point out
further work in this area.

\section{Cepheid relations at mean light}

The theoretical basis for the empirical Cepheid PL and PLC relations was first
outlined by Sandage (1958) as a consequence of the period mean density
relation for pulsating variable stars, the Stefan Boltzmann law and the
existence of a linear mass luminosity relation for Cepheids.
We outline this argument below.

The period mean density relation states that
\be
P{\overline{\rho}}^{1/2} = {\cal{Q}}
\ee
where $P$ is the Cepheid period, ${\cal{Q}}$ is a slowly varying function of
stellar parameters and ${\overline{\rho}}$, the mean density, satisfies
\be
{\overline{\rho}} \propto {\cal{M}} {\cal{R}}^{-3}
\ee
where ${\cal{M}}$ is the total mass and ${\cal{R}}$ is the radius of the star.
The Stefan Boltzmann law, however, states that
\be
L \, \propto \, {\cal{R}_{\mathrm{\ti{eq}}}}^2 \, T_{\mathrm{\ti{e}}}^4
\ee
where $L$ and ${\cal{R}_{\mathrm{\ti{eq}}}}$ are the equilibrium
luminosity and radius respectively,
and $T_e$ is the effective temperature. For Cepheids, we assume that the
equlibrium $L$ and ${\cal{R}_{\mathrm{\ti{eq}}}}$ are close to their
average values over a pulsational cycle. It then follows that
\be
{\cal{R}}^{3/2} \propto L^{3/4} \, T^{-3}_{\mathrm{\ti{e}}}
\ee

Substituting equations (2) and (4) into equation (1), the
period mean density relation, and taking logarithms we
obtain
\be
\log P + {1\over2}\log {\cal{M}} - {3\over4}\log L
+ 3\log T_{\mathrm{\ti{e}}} = \log{\cal{Q}}
\ee

If we assume that there exists a power law mass-luminosity relation
for Cepheids,
\be
\log L = \alpha \log {\cal{M}} + \beta
\ee
where $\alpha$ and $\beta$ are constants, then equation (5) may be reduced
to
\be
\log P +  ({\frac{1}{2\alpha}} - {\frac{3}{4}})\log L
+ 3 \log T_{\mathrm{\ti{e}}} = \log {\cal{Q}}
+ {\frac{\beta}{2\alpha}}
\ee

Equation (7) is the theoretical basis for the Cepheid PLC relation, and
its projection onto the plane of mean magnitude and period gives the PL
relation -- examples of which are given in Figure 4 of Madore and
Freedman (1991). It can be seen that there is a scatter about the
regression line of mean magnitude on period. That is,
for a given period
there is a range of mean magnitudes.
This dispersion results from a combination of several
different factors --
including the effect of reddening, which may not be
properly corrected, and observational errors in the
apparent visual magnitudes. Another factor
contributing to the dispersion is intrinsic, however,
and is caused by the finite width of
the instability strip in the HR diagram. A Cepheid can be
brighter than that luminosity given by the regression line by
having a hotter
surface temperature than a Cepheid of the same period but with a
mean magnitude given exactly by the regression line. Similarly,
a Cepheid of given period can be dimmer than the regression line
by having a cooler surface temperature. This is completely in
accord with equation (7). Hence the possible
range of Cepheid surface temperatures
(ie. the range of $T_{\mathrm{\ti{e}}}$ in equation (7) at a given period),
as well as the other factors mentioned above, leads to
the scatter in the Cepheid PL relation (Sandage 1958). The possible
range of Cepheid surface temperatures is determined by the width
of the instability strip - the locus of $L$ and $T$ points in
the HR diagram in which pulsation occurs. It is also clear that
a smaller range of $T_{\mathrm{\ti{e}}}$ in equation (7) will lead to a
tighter correlation between $P$, $\log L$, and $\log T$. We
discuss instances in which this is the case in the next section.

There has been considerable debate in the literature over
the exact cause of the scatter in the Cepheid PL relation.
Some authors claim that observational errors and an
incorrect allowance for reddening combined with a narrow, but non zero,
intrinsic width of the instability strip mean that it is difficult to
disentangle the effects of reddening and intrinsic
temperature variations (Madore and Freedman 1991; Clube and Dawe 1980;
Stift 1982, 1990).
Others maintain that observations and estimates of reddening and
colour excess are accurate enough to imply that the
scatter in the PL relation is due to intrinsic temperature
variations (MWF; CC; Feast and Walker 1987).
The arguments presented in these papers and the work of
Laney and Stobie (1986) who showed the existence of a significant
colour term in the infrared PLC relation for LMC Cepheids provide
ample support for this latter view.

\section{The Cepheid at Maximum light}

At maximum light, the period mean density law is still valid.
We assume, moreover, that the Stefan Boltzmann law still applies,
\be
L_{\mathrm{\ti{max}}} \, \propto \,
{\cal {R}}_{\mathrm{\ti{max}}}^2 \, T_{\mathrm{\ti{max}}}^4
\ee
where $L_{\mathrm{\ti{max}}}$ is the maximum luminosity,
${\cal {R}}_{\mathrm{\ti{max}}}$ is the
radius at light maximum and $T_{\mathrm{\ti{max}}}$ is the photospheric
temperature at maximum light. In the optical and
ultraviolet most of the light variations
in Cepheids are caused by temperature fluctuations, not radius
variations (c.f. Cox 1974, McGonegal et al 1982) and in any case maximum light
occurs when the star is expanding through its equilibrium
radius (Cox 1974; SKM), so we can also assume,
\be
{\cal {R}}_{\mathrm{\ti{max}}} \approx {\cal {R}}
\ee
where ${\cal {R}}$ is the equilibrium radius of the star.
However, even at longer wavelengths, the relative radius fluctuations
are no more than five to ten percent (Cox 1974) and
equation (9) will still be approximately valid. Thus our physical
derivation would still be applicable in the infrared. Therefore,
using equation (8) and (9) we obtain,
\be
{\cal {R}}^{3/2} \, \propto \, L_{\mathrm{\ti{max}}}^{3/4} \,
T_{\mathrm{\ti{max}}}^{-3}
\ee

Substituting equation (10) into the period mean density law,
taking logarithms and assuming equation (6),  yields
\be
\log P + {\frac{1}{2\alpha}} (\log L - \beta)
- {\frac{3}{4}} \log L_{\mathrm{\ti{max}}}
+ 3\log T_{\mathrm{\ti{max}}} = \log {\cal{Q}}
\ee

Equation (11) suggests the existence of a relation
between period, mean luminosity, maximum luminosity and maximum temperature.
The appearance of the $T_{\mathrm{\ti{max}}}$ term in equation (11) is
important because at maximum light, the range of Cepheid photospheric
temperatures is smaller (about $600$ K) than at mean light
(about $1000$ K), as was shown in SKM. This is
because at maximum light the photosphere occurs at the base of the
hydrogen ionization zone, independent of the period of the Cepheid.
For typical Cepheid densities and temperatures, the hydrogen ionization
zone occurs at about $6200$ K, although very long period
Cepheids ($P > 40$ days) have such extended envelopes that
the photosphere is further out in the envelope. Consequently these
longer period Cepheids have at maximum light a range of
photospheric temperatures similar to that at mean light (SKM).
If one considers only Cepheids of shorter period ($P < 40$ days), however,
then one would expect the range of $T_{\mathrm{\ti{max}}}$ in equation (11)
to be smaller than the
range of $T$ in equation (7). Consequently one can expect that PL and PLC
relations based on maximum light may have smaller scatter than those
based on mean light.

Another advantage of this approach is the following.
The Stefan Boltzmann law applied at mean and maximum light
yields,
\be
L^{-3/4} \, T_{\mathrm{\ti{e}}}^3 \, = \,
L_{\mathrm{\ti{max}}}^{-3/4} \, T_{\mathrm{\ti{max}}}^3
\ee
which implies,
\be
- {\frac{3}{4}} \log L + 3 \log T_{\mathrm{\ti{e}}} =
- {\frac{3}{4}} \log L_{\mathrm{\ti{max}}}
+ 3 \log T_{\mathrm{\ti{max}}}
\ee
Substituting for $\log T_e$ from equation (7) into the above
equation yields equation (11). We are constructing a relationship
using the properties of the pulsation at two phase points
(mean and maximum) to obtain equation (11), rather than
using just the properties at mean light as in equation (7).
Hence Cepheid relations based on equation (11) incorporate more
information about the pulsation than their counterparts based on
equation (7).

We can rewrite equation (11) in the following form:
\begin{eqnarray}
\log P & + & {\frac{1}{2\alpha}} (\log L - \log L_{\mathrm{\ti{max}}} +
\log L_{\mathrm{\ti{max}}} - \beta)
- {\frac{3}{4}} \log L_{\mathrm{\ti{max}}} \nonumber \\
 & + & 3 \log T_{\mathrm{\ti{max}}} = \log {\cal {Q}}
\end{eqnarray}
which implies,
\begin{eqnarray}
\log P & + & {\frac{1}{2\alpha}} (\log L - \log L_{\mathrm{\ti{max}}})
+ ({\frac{1}{2\alpha}} - {\frac{3}{4}})
\log L_{\mathrm{\ti{max}}} \nonumber \\
 & + & 3 \log T_{\mathrm{\ti{max}}} = \log {\cal {Q}}
\end{eqnarray}

Equation (15) suggests the possibility of a period, maximium
light, semi-amplitude relationship.

We can convert equation (11) into a form which can be
more easily compared with observations by writing,
\be
\log L = 0.4 \, (c - M_{\mathrm{\ti{bol}}})
\ee
and
\be
\log L_{\mathrm{\ti{max}}} =
0.4 \, ( c - M_{\mathrm{\ti{bol}}})_{\mathrm{\ti{max}}}
\ee
where $M_{\mathrm{\ti{bol}}}$ denotes the absolute bolometric magnitude,
which is related to the absolute visual magnitude, $M_{\mathrm{\ti{v}}}$, by
\be
M_{\mathrm{\ti{v}}} = M_{\mathrm{\ti{bol}}} + \mathrm{BC}
\ee

We can convert the bolometric correction, $\mathrm{BC}$, and the
temperature to an observed colour using relations of the
form,
\be
\mathrm{BC} = a + b \log (B-V)_0
\ee
and
\be
\log T = x + y (B-V)_0
\ee
where $a, b, x$ and $y$ are constants and $(B-V)_0$ is the
dereddened colour.

Combining equations (15) - (20) we have
\begin{eqnarray}
\log P - {\frac{0.2}{\alpha}} M_{\mathrm{\ti{v}}}
 & + & 0.3 M_{\mathrm{\ti{vmax}}}
+ ({\frac{0.2}{\alpha}} b + 0.3b + 3y) (B-V)_0 \nonumber \\
 & = & \mathrm{constant}
\end{eqnarray}

Clearly one can also convert the period, maximum light, semi-amplitude relation
suggested by equation (15) into a similar form involving magnitudes and
colours.
Much as equation (7) is the theoretical justification for the Cepheid PLC and
PL relations
at mean light (Sandage 1958; Sandage and Tammann 1968), equation (11) is the
theoretical justification for Cepheid PLC and PL relations at
maximum light. Although Sandage and Tammann (1968) constructed a period
maximum light relation for Cepheids in the Large and Small
Magellanic Clouds, M31 and NGC 6822, to our knowledge the above
argument has never been given.

\section{Cepheids as distance indicators}

As we indicated in Sect. 1, our main motivation
in studying and extending
PL and PLC relations for Cepheids is their usefulness as primary distance
indicators. In this section we illustrate how one would apply our
new relations, calibrated with a sample of Cepheids of known distance, to
infer the distance of other Cepheids grouped in a more distant galaxy or
cluster.
We follow the statistical formalism and notation adopted in HS90 and HS94
in discussing optimal galaxy distance indicators
such as the Tully--Fisher or $D_{\mathrm{\ti{n}}}-\sigma$ relations.
In particular, we adopt the
standard statistical convention of denoting an estimator of a quantity
by a caret. In order to avoid a surfeit of confusing subscripts
we also drop the subscript `$v$' from the absolute and apparent
visual magnitude.

The basic relationship between the absolute magnitude, $M$, and
apparent magnitude, $m$, (assumed corrected for absorption) for an object
at distance $D$ Mpc is
\be
m = M + 5 \log D + 25
\ee

For Cepheids, these magnitudes are usually taken to be the mean value over
their pulsation cycle. Clearly, however, for any Cepheid we also have
\be
m_{\mathrm{\ti{max}}} = M_{\mathrm{\ti{max}}} + 5 \log D + 25
\ee

An obvious estimate of the distance, $D$, (or more correctly the log distance)
of the Cepheid is then
\be
\widehat{\log D} = 0.2 ( m - \hat M - 25 )
\ee
or
\be
\widehat{\log D} = 0.2 ( m_{\mathrm{\ti{max}}} - \hat M_{\mathrm{\ti{max}}}
- 25 )
\ee

Here $\hat M$ and $\hat M_{\mathrm{\ti{max}}}$ denote estimators of the
mean and
maximum absolute magnitude respectively. Examples of such estimators
include
\be
\hat M = a_{1} + b_{1} \log P
\ee
or
\be
\hat M = a_{2} + b_{2} \log P + c_{2} (B-V)_{0}
\ee
where $P$ and $(B-V)_{0}$ take their usual meanings and $a_{1}$,
$b_{1}$, $c_{1}$, $a_{2}$ and $b_{2}$ are
constants. The motivation for considering estimators of this form is clearly
the PL and PLC relations -- the theoretical basis for which has already
been described in the preceding sections.

The constant coefficients in equations (26) and (27)
are obtained by fitting such relations to the relevant observations --
i.e. to the {\em apparent\/} visual magnitude, period and colour -- for
Cepheids of known distance. For example, if the calibrating Cepheids
are all at distance, $D_{\mathrm{\ti{cal}}}$ Mpc,
then we may rewrite equation (26) as
\be
\hat m - 5 \log D_{\mathrm{\ti{cal}}} - 25 = a_{1} + b_{1} \log P
\ee

In other words the estimator of absolute visual magnitude given by
equation (26)
is equivalent to the estimator of {\em apparent\/} visual magnitude
given by equation (28).
That is, in order to best fit the intrinsic relations
involving mean and maximum absolute magnitudes,
we need only consider the corresponding relations involving
apparent magnitudes, provided all the calibrating stars are assumed
equidistant. Similar remarks apply to all the PLC--type relations discussed
in this paper.

The fitting procedure mentioned above is usually a maximum likelihood or
linear regression analysis, and requires a statistical model for the
relationship between the relevant variables. A suitable model for the PL
relation, for example, might be
\be
M = a + b \log P + \epsilon
\ee
where the errors, $\epsilon$, are taken to be normally distributed
with mean $0$ and
variance ${\sigma}^2$. Given such an error model, the estimator, $\hat M$,
given by equation (26)  with $a_{1} = a$ and $b_{1} = b$
corresponds to the expected absolute magnitude conditional upon log period,
and the values of $a$ and $b$ are in this case equal to the slope and
zero point respectively of the direct linear regression of $M$ on $\log P$.
An example of a more sophisticated model for the PLC relation is
given in the appendix of CC. In the appendix of this paper we describe the
statistical model adopted in the present analysis.

Given an appropriate statistical model it is straightforward to
determine the distribution, $p (\widehat{\log D} | D_{\mathrm{\ti{T}}})$, of
$\widehat{\log D}$ conditional upon the true Cepheid distance,
$D_{\mathrm{\ti{T}}}$.
One can then use the moments of this distribution as a suitable criterion
for comparing the properties of different estimators. For example, a
desirable property is that $\widehat{\log D}$ be {\em unbiased\/}. This
requires that $\widehat{\log D}$ satisfies the relation
\begin{eqnarray}
E (\widehat{\log D}|D_{\mathrm{\ti{T}}}) \, & = \, \int \widehat{\log D}
\, p(\widehat{\log D} | D_{\mathrm{\ti{T}}}) \, d \widehat{\log D}\\
                              & = \; \log D_{\mathrm{\ti{T}}} \nonumber
\end{eqnarray}
(c.f. eqn. [4] of HS94).

In other words $\widehat{\log D}$ will, on average, yield the true
log distance of the Cepheid, whatever that true distance is.
In a similar manner the {\em risk\/}, ${\cal{R}}$, of an estimator is defined
as the expected
value of the second moment of $p(\widehat{\log D} | D_{\mathrm{\ti{T}}})$,
i.e.
\be
{\cal{R}} \, = \, \int ( \widehat{\log D} - \log D_{\mathrm{\ti{T}}} ) ^{2}
\, p(\widehat{\log D} | D_{\mathrm{\ti{T}}}) \, d \widehat{\log D}
\ee
(c.f. eqn. [5] of HS94). Note that for an unbiased estimator the risk is
identically equal to the variance. In this study we will consider as optimal
log distance estimators which are unbiased and which have minimum risk or
variance -- a criterion which we discuss more fully in the appendix.

Generally, better physics should translate into better
statistics. For example, equation (27) reflects a more complete
physical model for $\hat M$ than that given by equation (26),
and we would therefore
expect that the distance estimator of equation (24)
would have a smaller variance if $\hat M$ is given by equation (27)
 rather than equation (26).
In fact, previous studies such as MWF and CC have indeed shown that
for LMC Cepheids, the introduction of a colour term offers a
significant reduction in scatter over a PL relation.

In this study we investigate the variance of log distance estimators
taking the form of equations (24) and (25), where our
models for $\hat M$ and $\hat M_{\mathrm{\ti{max}}}$ are prompted by
equations (11) and (15) respectively.

It follows trivially from equations (22) - (25) that for an unbiased estimator
${\cal{R}}$ is given by
\be
{\cal{R}} = 0.04 E({\hat M}_{\ti{*}} - M)^2 = 0.04
{\sigma}^2_{{\hat M}_{\ti{*}}}
\ee
where an asterisk denotes 'mean' or 'maximum' magnitude as appropriate.
The rms percentage distance error, $\Delta$, of ${\widehat {\log D}}$
is given by
\be
\Delta \simeq 20 \, {\mathrm{ln}} 10 \, {\sigma}_{{\hat M}_{\ti{*}}}
\ee

Specifically, in this paper we examine
the following estimators of mean or maximum absolute magnitude.
\be
\hat M = a + b \log P
\ee
\be
\hat M_{\mathrm{\ti{max}}} = a + b \log P
\ee
\be
\hat M = a + b \log P + c (B-V)
\ee
\be
\hat M_{\mathrm{\ti{max}}} = a + b \log P + c (B-V)_{\mathrm{\ti{max}}},
\ee
\be
\hat M_{\mathrm{\ti{max}}} = a + b \log P
+ c (B-V)_{\mathrm{\ti{max}}}
+ d ( M_{\mathrm{\ti{mean}}} - M_{\mathrm{\ti{max}}} )
\ee
\be
\hat M_{\mathrm{\ti{max}}} = a + b \log P + c (B-V)_{\mathrm{\ti{max}}}
+ d M_{\mathrm{\ti{mean}}},
\ee

Of course the constants $a$, $b$, $c$ and $d$ are different in each case.
Again, to avoid a surfeit of subscripts we have dropped the subscript
`0' from $(B-V)$ and $(B-V)_{\mathrm{\ti{max}}}$. Henceforth, unless we
state otherwise, all colours will be assumed to be corrected for reddening.
Equations (34) and (36) are the standard Cepheid
PL and PLC relations at mean light. Equation (35) is a
Cepheid PL(max) relation, as described in Sandage (1968).
Equations (37), (38) and (39) are new and
are based upon equations (11) and (15).

Note that the semi-amplitude term on the right hand side of equation
(38) involves the {\em true\/} value of both $M_{\mathrm{\ti{mean}}}$ and
$M_{\mathrm{\ti{max}}}$, neither of which is directly observable
(otherwise we would have no need to estimate them!). This presents
no difficulty, since $M_{\mathrm{\ti{mean}}} - M_{\mathrm{\ti{max}}}$
may be re-written as
$m_{\mathrm{\ti{mean}}} - m_{\mathrm{\ti{max}}}$, i.e. the difference
between the mean and maximum
{\em apparent\/} magnitude, a quantity which is readily observable. Thus,
\be
{\hat M}_{\mathrm{\ti{max}}} = a + b \log P +
c (B-V)_{\mathrm{\ti{max}}} +
d(m_{\mathrm{\ti{mean}}} - m_{\mathrm{\ti{max}}})
\ee

The
presence of $M_{\mathrm{\ti{mean}}}$ on the right hand side of equation (39),
however, cannot be dealt with in this way. In order to overcome the
obvious problem that $M_{\mathrm{\ti{mean}}}$ is not directly
observable, and in addition to avoid fitting a relation which is
distance degenerate, we replace equation (39) by
\be{\hat M}^{'}_{\mathrm{\ti{max}}} = a + b \log P
+ c (B-V)_{\mathrm{\ti{max}}}
+ d (M_{\mathrm{\ti{mean}}} - < M_{\mathrm{\ti{mean}}}>)
\ee
where $<M_{\mathrm{\ti{mean}}}>$ denotes the sample averaged absolute
magnitude at mean light of a group of equidistant Cepheids (e.g. in a
distant galaxy or cluster, the
distance of which we wish to estimate).
Of course the zero point, $a$, of equation (41) will differ from that in
equation (39). Note that
$M_{\mathrm{\ti{mean}}} - < M_{\mathrm{\ti{mean}}}>$ is a distance independent
quantity (as is $M_{\mathrm{\ti{mean}}}$) but {\em is} directly observable,
being equal simply to
$m_{\mathrm{\ti{mean}}} - < m_{\mathrm{\ti{mean}}} >$.
Note also,
however, that the distribution of $<M_{\mathrm{\ti{mean}}}>$ depends upon the
sample size of Cepheids, and in particular, that one could not sensibly
apply equation (41) to estimate the maximum absolute magnitude of an
individual field Cepheid.

Some algebra easily establishes that the variance of
${\hat M}^{'}_{\mathrm{\ti{max}}}$ is given by the sum of the variance of
${\hat M}_{\mathrm{\ti{max}}}$ in equation (39) and the variance of
$<M_{\mathrm{\ti{mean}}}>$, i.e.
\be
{\sigma}^2_{{\hat M}^{'}_{\mathrm{\ti{max}}}} =
{\sigma}^2_{{\hat M}_{\mathrm{\ti{max}}}} +
{\sigma}^2_{\mathrm{\ti{M}}}/n
\ee
where ${\sigma}_{\mathrm{\ti{M}}}$ is the dispersion of the intrinsic
distribution of magnitudes at mean light (which of course is not known
a priori,
but which is estimated from the LMC calibrating Cepheids) and $n$ is the
number of observed Cepheids in the more distant galaxy or cluster. The risk,
${\cal{R}}$, of the corresponding log distance estimator is given by
${\cal{R}} = 0.04 {\sigma}^2_{{\hat M}^{'}_{\mathrm{\ti{max}}}}$, in
accordance with equation (32) above.

Finally in this section we should note that the above analysis takes no account
of the metallicity dependence of our fitted relations. Although the
Cepheid PL relations (34) and (35) are insensitive to
composition differences, this will not be the case for the remaining
PLC-type relations. One can overcome this problem as follows.
One fits equations (36) - (39) to the LMC data and then adjusts the
coefficients of the fit (using eg. Table (b1) in CC)
to give the PLC relation of a normal metal abundance Cepheid. The zero point
of this corrected relation can then be
then calibrated using Galactic Cepheids of known distance.
CC use the results of Iben and Tuggle (1975),
Becker, Iben and Tuggle (1977) and Bell and Gustafsson (1978) to
obtain their Table (b1). The composition dependendence of
the Cepheid mass luminosity law is part of the reason for
the metallicity sensitivity of the PLC.
It is not known
exactly how this will affect e.g. the PLC(max) relation.,
Since Feast and Walker (1987) suggest that the metal deficiency
of LMC Cepheids is slight, however, (1.4 times less than solar),
it seems likely that the
difference between the values of $b$ and $c$ in equation (37)
and those appropriate for a metal normal Cepheid will be small.
Because our primary intention in this paper is to establish how
the use of Cepheids at mean and maximum light affects the
{\em dispersion\/} of our distance estimators, we leave a more precise
zero point calibration of relations (36) to (39) -- accounting for
metallicity dependence as described above -- for subsequent papers.

\section{LMC Data}

In order to test our assertion of the existence of
Cepheid relations of the form suggested by equations (7), (11) and (15), we
use multicolour photoelectric observations of LMC Cepheids
taken by Martin and Warren (1979) and discussed in MWF. These
observations were taken in the BVI system.
The list of stars used in this analysis is given in Table 1,
together with their period in days. Note that some of the stars
in the original data set presented in Martin, Warren and
Feast (1979) were omitted because there were not enough
observed points to obtain accurate mean and maximum
magnitudes. These excluded stars are shown in Table 2.
Rejecting these stars
resulted in the total sample of 39 stars shown in Table 1.
MWF and Feast (1984) have discussed the small dispersion in
reddenings for their LMC Cepheids: hence we adopt a constant value of
$E(B-V)=0.1$ given by Madore and Freedman (1991).
We take $R = A_V/E_{\mathrm{\ti{(B-V)}}}$
to be 3.3, again following MWF. Since the only data set available to us was
in magnitudes and not intensity fluxes, our PL and PLC
relations are determined in terms of magnitudes.

Figure 1 shows a plot of period vs. semi-amplitude for
all 39 stars in our sample. It clearly shows a group of stars
at long period which are well separated from the
other stars in the sample. Rather than postulate why this is so,
we will present results with and without this set of stars.
These stars are presented in Table 3, marked by an asterix in the
third column.
Recall that in Sect. 3 and in SKM it was suggested that
at maximum light Cepheids have a smaller range of surface temperatures
than at mean light, but that for periods $P \geq 40$ days this
distinction is not found. Since the range of surface temperatures
also introduces scatter, we will consider a third subset of the
data consisting only of those stars with periods $P < 40$ days.
Stars with periods $P > 40$ days are
denoted by a `+' in Table 3.
Moreover, note that the
disjoint group of stars listed in Table 3 all have periods $P > 40$, and so
are also rejected from this third sample. We therefore consider three data
sets, with 39, 35 and 31 stars respectively.
\begin{figure}
  \centerline{\psfig{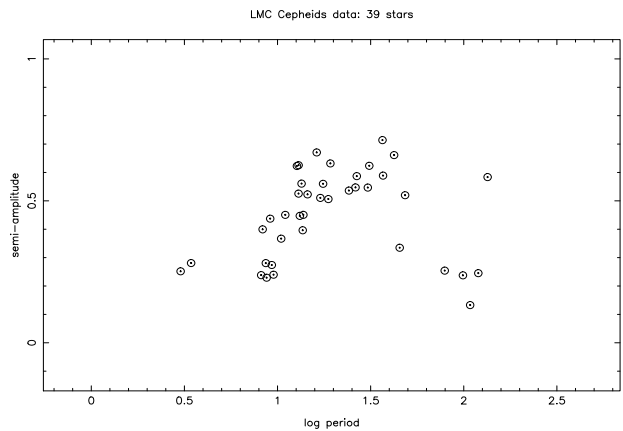}}
  \caption{log period plotted versus semi-amplitude for the 39 LMC
 Cepheids in our calibrating sample.}
\end{figure}
In the appendix, we introduce a multivariate normal model to describe the
joint distribution of the six relevant variables in the Cepheid data:
maximum magnitude, mean magnitude, maximum colour, mean colour, log period
and semi-amplitude.
We {\em require\/} this joint distribution to be normal in order
to make a strictly valid application in Sect. 6 of our hypothesis tests that
the
variance of distance estimators based on
maximum light is significantly smaller than for those based upon mean light.

We can test the normality of the distribution of each of these variables in
several ways. First, we plot a sample cumulative distribution function
(CDF) and compare this with the theoretical CDF of a normal distribution
with mean and dispersion equal to the sample mean and dispersion of that
variable. Figures 2, 3 and 4 show selected results obtained for the samples of
31, 35 and 39 stars respectively. The agreement with the theoretical normal
curves is generally quite good, although is somewhat worse for the
distribution of semi-amplitude than for the other variables.
\begin{figure}
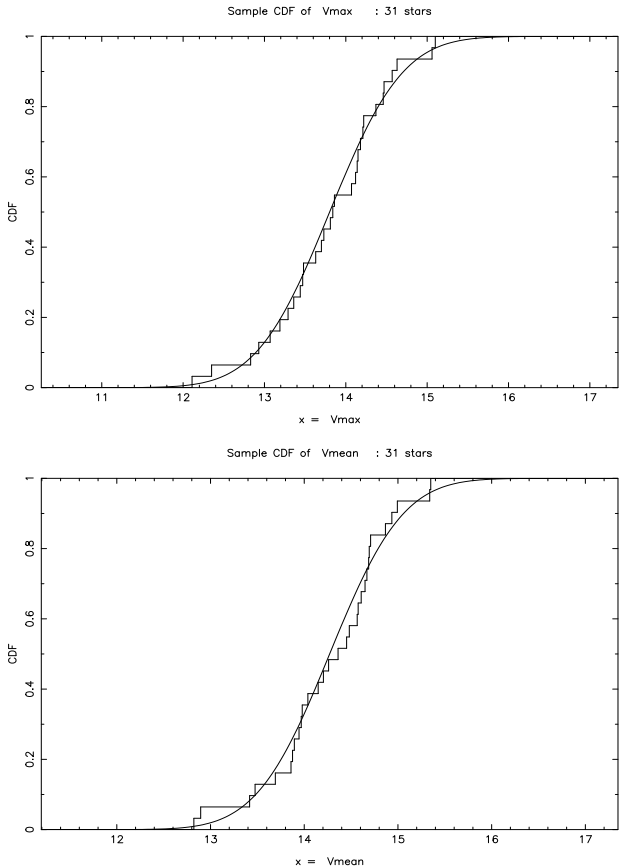

  \centerline{\psfig{file=figure2.ps.1,width=3.5in,height=2.3in,angle=-90}}
  \centerline{\psfig{file=figure2.ps.2,width=3.5in,height=2.3in,angle=-90}}
  \caption{Comparison of sample CDF curve with the normal curve of equal
 mean and dispersion. The examples shown are for mean and maximum
 apparent magnitude, for the subsample of 31 stars.}
\end{figure}
\begin{figure}
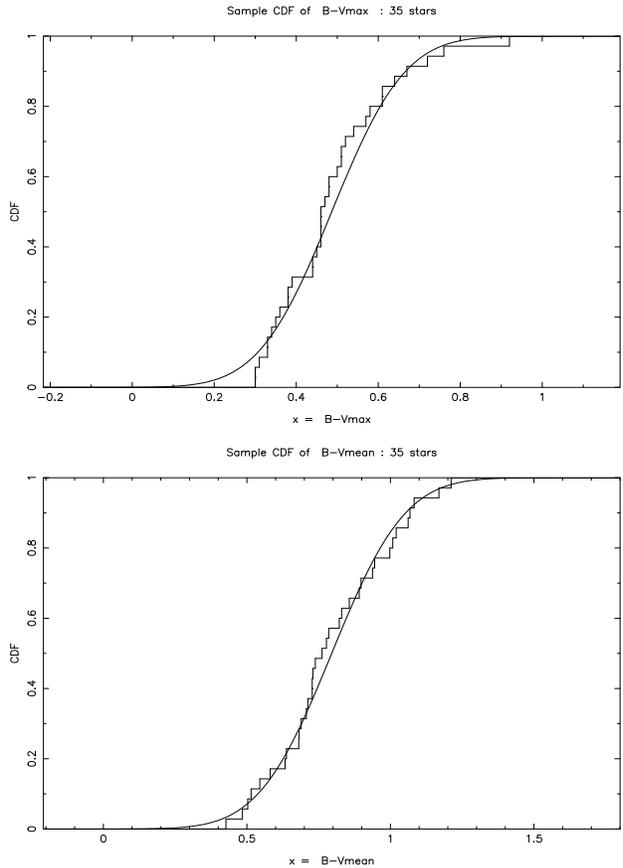

  \centerline{\psfig{file=figure3.ps.1,width=3.5in,height=2.3in,angle=-90}}
  \centerline{\psfig{file=figure3.ps.2,width=3.5in,height=2.3in,angle=-90}}
  \caption{Comparison of sample CDF curve with the normal curve of equal
 mean and dispersion. The examples shown are for mean and maximum
 colour, for the subsample of 35 stars.}
\end{figure}
\begin{figure}
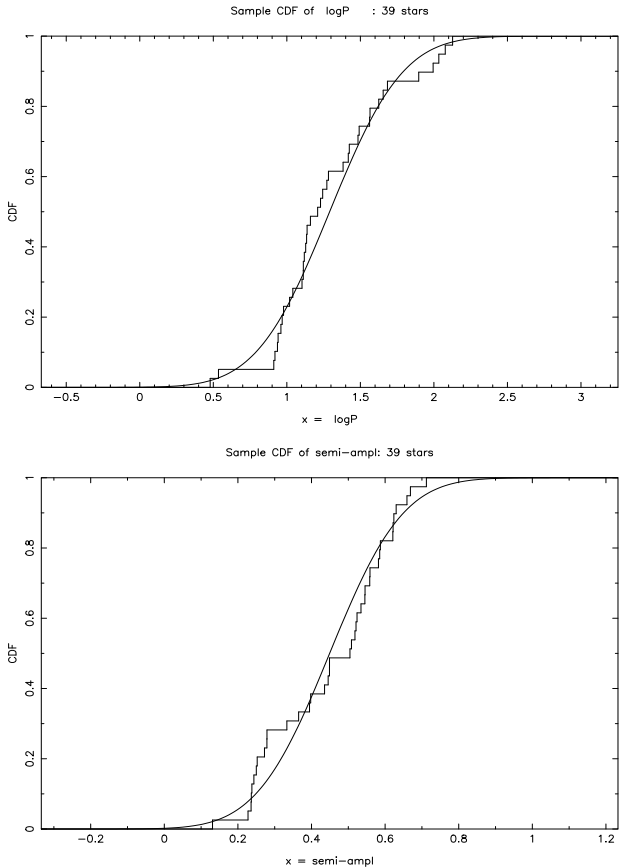

  \centerline{\psfig{file=figure4.ps.1,width=3.5in,height=2.3in,angle=-90}}
  \centerline{\psfig{file=figure4.ps.2,width=3.5in,height=2.3in,angle=-90}}
  \caption{Comparison of sample CDF curve with the normal curve of equal
 mean and dispersion. The examples shown are for log period and
 semi-amplitude, for the full sample of 39 stars.}
\end{figure}
To quantify this
agreement we apply the Kolmogorov--Smirnov (KS) test (c.f. Kendall \& Stuart,
1963), for which the test
statistic, $D_{\mathrm{\ti{n}}}$, is defined as the maximum absolute deviation
between the sample and model CDF curves. Table 4 lists
$D_{\mathrm{\ti{obs}}}$, the observed value of the test
statistic, for each of the six relevant observables and using each of
the three data sets under consideration,
together with the corresponding significance of the KS test -- i.e. the
probability that $D_{\mathrm{\ti{n}}} > D_{\mathrm{\ti{obs}}}$ under the
null hypothesis
that the sample CDF is drawn from the modelled theoretical distribution.

It is clear from these
results that, on the basis of the KS test, there is no strong evidence to
reject the hypothesis that the six variables are normally distributed --
although it is also clear that the hypothesis is much more strongly accepted
for the restricted data set of 31 stars than for the full sample.

The main advantage of the KS test is its robustness, however, and it
would seem
prudent to apply more powerful tests of normality to confirm the validity of
our assumptions. We next calculated the sample skewness, ${\cal{S}}$,
and kurtosis, ${\cal{K}}$, for each variable, $x$, defined as
\be
{\cal{S}} \, = \, {\frac{1}{N}} \sum_{j=1}^{N}
{ ( {\frac{x_{\mathrm{\ti{j}}} - \overline{x}}{{\hat{\sigma}}}} ) }^3
\ee
and
\be
{\cal{K}} \, = \, {\frac{1}{N}} \sum_{j=1}^{N}
{ ( {\frac{x_{\mathrm{\ti{j}}} - \overline{x}}{{\hat{\sigma}}}}
) }^4 \, - \, 3
\ee
where $\overline{x}$ and ${\hat{\sigma}}$ denote the sample mean and standard
deviation of the sample of size, $N$, respectively.

The intrinsic skewness and kurtosis should both be indentically zero for
a normal distribution. The sampled value of each
statistic will fluctuate around zero, however, and the variance of the
sampling distribution depends upon the sample size, $N$.
Under the null hypothesis of a normal variable, the variance of the
sample skewness and kurtosis is approximately equal to $15/N$ and $96/N$
respectively (c.f. Kendall and Stuart, 1963).

Table 5 lists the sample skewness and kurtosis, expressed in terms of the
number of standard deviations under the null hypothesis of a normal
distribution, for each of our six Cepheid observables and for the three
different sample sizes considered. From these results we see that the
sample of 31 stars shows no significant evidence of non-zero skewness and
kurtosis -- the largest sampled skewness being $\sim 0.72 \sigma$ for
mean apparent magnitude. In the larger samples, however, there is some
indication of significant skewness above the one $\sigma$ level for
mean magnitude and maximum colour.

The final test which we apply is based upon Shapiro and Wilks' $W$
statistic (Royston, 1982). This is a considerably more powerful
test and uses the properties of the {\em order statistics\/} of a normal
distribution -- i.e. the properties of a sample arranged in ascending or
descending order. (c.f. David, 1981; Hendry, O'Dell and Collier Cameron, 1993).
The test statistic is given by
\be
W \, = \, \frac{{ ( \sum a_{\mathrm{\ti{j}}}
x_{\mathrm{\ti{j}}} ) }^2}
{\sum ( x_{\mathrm{\ti{j}}} - \overline{x} )^2}
\ee
where $\overline{x}$ denotes the sample mean and
the set of normalising weights,
$a_{\mathrm{\ti{j}}}$, depend upon the sample size, $N$.

The mean value of the statistic
under the null hypothesis of a sample drawn from a normal distribution
is unity.

Table 6 shows the values of the $W$ statistic and the calculated significance
of the test in each relevant case. From these results we see that
our sample of 31 stars satisfies well the assumption of normality: only
for the observed semi-amplitudes is there any evidence, at the $\sim 5 \%$
level,
for significant deviation from a normal distribution. In the larger samples,
however, the validity of the normal assumption is more
marginal. In fact for our full sample of 39 stars only for the mean
colour would the null hypothesis
be accepted at the $15 \%$ level, and for the maximum colour it would be quite
strongly rejected, at a level of $\sim 0.01 \%$. Clearly, then, we can
safely apply hypothesis tests based on normality to the restricted sample
of 31 stars, but must be somewhat more cautious in drawing conclusions from
their application to the larger samples. We will comment further upon this
point in Sect. 6 and Sect. 7.

Finally, it is important to note that -- in modelling the joint
distribution of the six relevant physical variables as multivariate
normal -- we do not make any explicit
assumptions concerning the nature of the scatter in their observed
distribution. In particular we do not attempt to
separate intrinsic scatter and measurement error in their observed
distribution.
Undoubtedly some component of the variance in our fitted distance relations
will be due to observational errors and related effects such as line of sight
spread in the true distance of the calibrating stars -- both effects which
are, at least in principle, removable. Our point is that the relative
contribution of intrinsic scatter and observational errors is unlikely to
differ dramatically in each of our distance estimators. Hence
our conclusions concerning the {\em relative\/} reduction
in variance resulting from the use of magnitudes and colours at maximum light
will not be
significantly changed -- even if the absolute value of the variance could
be reduced in both cases by the use of more precise observations
(observations which -- notwithstanding these remarks -- are now available
from HST and 8m class terrestrial telescopes, and from the use of observations
at redder wavelengths).

\section{Results and Discussion}

In this section we present the results of fitting the coefficients of the
relations defined in equations
(34) - (39) by a multilinear regression model as described in the appendix.
Each relation has been fitted using the three LMC samples -- of 31, 35 and 39
stars respectively. The relations have been converted to absolute magnitudes
at mean and maximum light assuming an LMC distance modulus of 18.5 mag for the
LMC (c.f. Madore and Freedman, 1991; Pierce et al., 1994). Tables 7 and 8
present
fitted PL and PLC relations respectively at mean and maximum light, and also
a period, luminosity, semi-amplitude (PLA)
relation - again fitted at both mean and maximum light.
Table 9 presents the results of fitting a period, maximum luminosity, mean
luminosity, maximum colour (PLL'C) and period, maximum luminosity,
semi-amplitude, maximum colour (PLAC) relation.

The table headings indicate the form of the fitted
relation and the variables used in the fit. The next rows indicate the
values of the regression coefficients obtained. The final two columns
of each table give the
variance of ${\hat M}_*$ derived from the linear regression (where, as before,
a star indicates mean or maximum light as appropriate) and the percentage rms
error of the corresponding distance estimator, calculated from equation
(33) - with the exception of the PLL'C relation in table 9, the distance error
dispersion of which we discuss separately. Next to the fitted regression
coefficients for each relation we give their computed standard errors. For the
coefficients of the relevant physical variables in each fit, we also indicate
the computed probability of obtaining a sample regression coefficient larger
in modulus than the estimated value, under the null hypothesis that the
true regression coefficient is identically zero, applying the t test described
in the appendix. This provides a clear and
useful indication of the relative importance of the different independent
variables in each relation. Finally, for the relations with two or more
independent variables, we also present beneath each table
the results of a second significance test involving the
partial sample multiple correlation coefficient (SMCC) of each regression fit,
as described in the appendix. This test provides a direct measure
of the significance of including the final independent variable in reducing the
overall dispersion of the fit. We list the partial SMCC for each fit
and the probability that the value of the test statistic $W$
be greater than its observed value under the
null hypothesis that the true partial SMCC be identically zero: i.e. that the
final independent variable makes no contribution to the reduction of the
scatter in the relation.

\subsection{PL relations}

Table 7 presents our results for equations (34) and (35), the
PL relations at mean and maximum light. Note that there is hardly any
difference in the variance of the fit, or the dispersion of the corresponding
distance estimator, between mean and maximum light for each of the three
samples.
As one would naturally expect, the regression
coefficient of $\log P$ is in all cases highly significant: cepheid
magnitudes at mean or maximum light are not well described by a constant.
The coefficient of
$\log P$ in each case varies from about -2.0 to -2.5, and is larger in modulus
for the maximum light relation for each sample size, with comparable
standard error -- although the difference
is always within 2 $\sigma$.
This range is somewhat different from existing PL
relations for the LMC -- for example -2.88 (MF), -2.69 (CC) and a range
from -2.59 to -2.90 (MWF). Table 3 of MWF lists the data set used
in their work: as we noted in Sect. 5, our data set is a subset of this sample.
The results of table 7 indicate, therefore, that the coefficient of $\log P$
in a linear least squares fit to a subset of the MWF data set differs from the
value obtained when the entire data set is used. This does reinforce the
fact that PL relations are inherently statistical in nature and care must be
taken in comparing the results of studies which sample different period
ranges -- a point also recently addressed in Pierce et al., 1994.
In the present context, this means that we
can usefully compare the dispersion of different distance relations obtained
for the {\em same} sample -- of 31, 35 or 39 stars -- but must be more cautious
in comparing results for a given relation across the three samples,
particularly in the light of the normality test results described in Sect. 5.

\subsection{PLC and PLA relations}

Table 8 presents our results for PLC relations, based on equations (36) and
(37), at mean and
maximum light. We can see that there is an appreciable reduction in
the dispersion of each relation, and for all three samples, compared with
the corresponding PL relations. This conclusion is borne out in a
quantitative manner in several ways. First note that the t test applied to the
regression coefficients strongly rejects the null hypothesis of a zero
regression coefficient for colour in all cases,
although it is interesting to note that the significance of a non-zero
regression coefficient of $\log P$ is still considerably higher than that
for colour, suggesting that the greater contribution to the reduction of
scatter is coming from the PL relation. Notwithstanding this, the results of
the $W$ test applied to the partial SMCCs for each relation do confirm that
the addition of colour significantly reduces the scatter compared with
the PL relation.
In all cases the null hypothesis of no reduction in scatter is rejected
strongly, with a probability of acceptance of
$4.9 \times 10^{-4}$ or less. Hence, our
application of this hypothesis test reconfirm the results of
MWF and CC who established the existence of a colour
term for LMC Cepheids. The coefficients of
$\log P$ and $(B-V)_{\mathrm{\ti{mean}}}$ which we have
obtained are in all cases
within one standard error of the values given by MWF and CC.

We can see from table 8 that the PLC relation at maximum light leads to
a distance estimator with about $10 \%$ less dispersion
than its counterpart at mean light for the three samples. This result is
further supported
by the fact the partial SMCC test more strongly rejects the null hypothesis
for the PLC relation at maximum light than at mean light. Applying an $F$ test
to determine the significance of this reduction in dispersion, we find a
significance of 0.27, 0.24 and 0.34, for the samples of 31, 35 and 39 stars
respectively. Thus, our sample of calibrating Cepheids is still too small
to confirm at a high significance level the improvement of the PLC
relation at maximum light.

Figures 5 and 6 show plots of period against $(B-V)$ colour at mean and maximum
light respectively. From these plots one can conclude that there is only a
very small reduction in the width of the instability strip at maximum light
for our LMC sample: a fact which may well account for the inferred reduction
in scatter of only $10 \%$ in the maximum light PLC relation.
Moreover, it can be seen from figure 3 that
the maximum $(B-V)$ colour is essentially independent of period for
$\log P < 1.5$ (c.f. SKM). Figure
7 of SKM, however, illustrates that the temperature range of Galactic
Cepheids at maximum light is only about $600$ K.
This suggests that a PLC relation at maximum light constructed for
Galactic Cepheids may lead to bigger reduction in scatter compared to that
at mean light.
\begin{figure}
  \centerline{\psfig{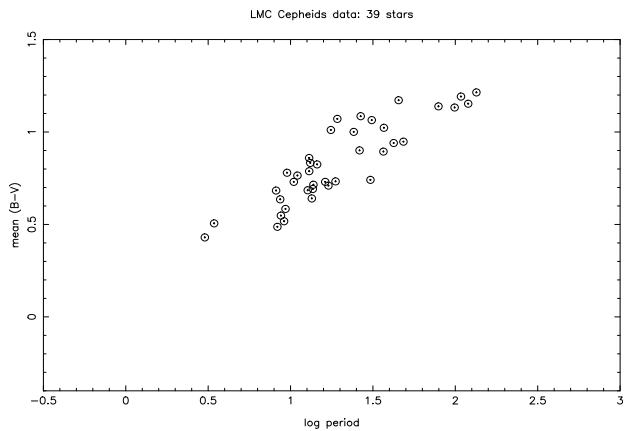}}
  \caption{log period plotted against unreddened $B-V$ colour at mean light,
  for the full sample of 39 LMC stars}
\end{figure}
\begin{figure}
  \centerline{\psfig{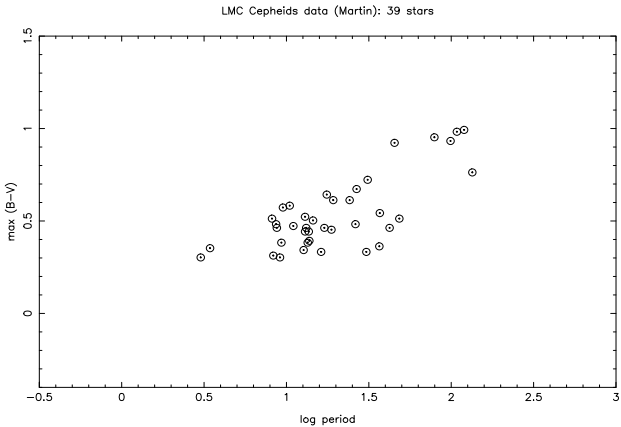}}
  \caption{log period plotted against unreddened $B-V$ colour at
  maximum light, for the full sample of 39 LMC stars}
\end{figure}

Table 8 also presents the results of fitting PLA relations to the LMC data,
which represent a projection of the PLAC relation suggested by equation
(38) and considered below.
In all cases it was found that the addition of a semi-amplitude term did
not significantly reduce the dispersion of the relation. The partial
SMCC test accepted the null hypothesis at a level of more than $10 \%$
and the regression coefficient of semi-amplitude was not significantly
different from zero in any case. What is interesting to note, however,
is that the PLA relation appeared to fare best when applied to the full
sample of 39 stars in the sense that, in this case, the null hypothesis
of the partial SMCC test is accepted least strongly. This is perhaps not too
surprising since the 39 star sample includes the group of four stars which,
from figure 1, clearly have unusual amplitudes - and one might expect that a
relation involving an amplitude term would be most readily able to accommodate
such stars. It would be unwise to attach too much importance to this
result, however, since we have shown in Sect. 5 that the semi-amplitudes
satisfy the assumption of normality least accurately, which may in turn
adversely affect the results of any hypothesis test based on this assumption.

\subsection{PLL'C and PLAC relations}

Table 9 presents the results of our fits to the PLL'C relation, in the
modified form of equation (41), and PLAC relation given by equation (38).
Firstly note that the PLAC relation does not offer a significant reduction in
dispersion over the PLC relation at mean or maximum light in the samples of
31 or 35 stars, consistent with the results of the previous section.
There {\em is\/} a significant improvement in the relation for the 39 star
sample, however. The partial SMCC null hypothesis is rejected at the
$3 \%$ level and the regression coefficient of semi-amplitude is
consistent with zero at a similarly small probability.
This would again seem to be
due to the inclusion of the group of stars of unusual amplitude in our
sample. Notwithstanding our earlier note of caution about the consistency
of the semi-amplitude distribution with a normal distribution, we
nevertheless believe this result to indicate that a semi-amplitude
term has an important role to play in improving Cepheid PLC relations --
not as a replacement for, but in addition to, a colour term -- when stars
of unusual amplitude for their period are considered.

The PLL'C relation, on the other hand, is seen to have a considerably reduced
dispersion -- by a factor of two or more -- for all three samples. The
partial SMCC null hypothesis is now very strongly rejected and the
regression coefficient of
$M_{\mathrm{\ti{mean}}} - < M_{\mathrm{\ti{mean}}}>$
differs from zero with a very high significance and small standard error.

The outstanding difficulty with the PLL'C relation lies with how one can
convert it
into a useful distance estimator. The basic problem is that absolute
magnitude at mean and maximum light are related to their apparent
magnitude couterparts via an identical function of distance, so that
equation (39) may be distance--degenerate: i.e. it yields essentially
no distance information if the coefficient of $M_{\mathrm{\ti{mean}}}$ is
close to unity.
As we saw in Sect. 4, we have attempted to overcome this
problem by introducing the sample averaged absolute magnitude at mean light
in a group of distant Cepheids -- effectively turning the
$M_{\mathrm{\ti{mean}}}$ term in equation (39) into a quantity which is both
distance independent and directly observable -- but at the cost of
increasing the dispersion of our distance estimator due to the sampling
variance of $< M_{\mathrm{\ti{mean}}}>$, and thus restricting the
application of our PLL'C relation to a sufficiently large group of distant
Cepheids.

In figure 7 we plot the rms percentage error dispersion, $\Delta^*$, of the
corresponding distance estimator, divided by the percentage distance
error of the PLC relation at maximum light, as a function of $n$, the number
of distant Cepheids observed. We use the values of
${\sigma}^2_{{\hat M}_{\mathrm{\ti{max}}}}$ and
${\sigma}^2_{\mathrm{\ti{M}}}$ derived from the LMC sample of 31 stars, and
compute $\Delta^*$
using equations (33) and (42). We can see from figure 8 that $\Delta^*$
falls off quite slowly with $n$. Moreover, we see that for small samples of 10
or fewer Cepheids $\Delta^*$ is greater than unity -- i.e. the
resultant distance error dispersion of the PLL'C relation is in fact
{\em larger} than that for the PLC relation. This is because the reduction
in the variance of ${{\hat M}_{\mathrm{\ti{max}}}}$ is more than
cancelled out by the large sample variance of $< M_{\mathrm{\ti{mean}}}>$.
For larger values of $n$, however, we do see an appreciable reduction
in the dispersion of the distance estimator, falling to $\sim 63 \%$ for
$n = 30$ and to $\sim 53 \%$ for $n = 40$. It seems, therefore, that
our modified PLL'C can significantly improve upon the PLC relation for
fairly large but realistic sample sizes of distant Cepheids. Similar
results are obtained for the samples of 35 and 39 Cepheids.
\begin{figure}
  \centerline{\psfig{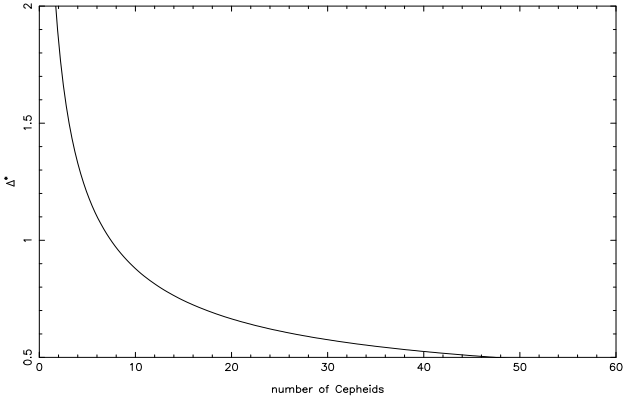}}
  \caption{rms percentage distance error, $\Delta^*$, derived from the
  PLL'C relation -- divided by the percentage distance error of the
  PLC relation at maximum light -- as a function of $n$, the number of
  observed Cepheids.}
\end{figure}

Of course we should note here that we are neglecting the effect of
sampling variance on the determination of the distribution parameters in
our LMC calibrating data: clearly with a calibrating sample of less than
40 stars this effect cannot be considered negligible. It will have no
bearing upon our comparison of the relative dispersion of any of our
estimators, however, since it will affect every relation in precisely the
same way. Moreover, its effect on the calibrating sample is in principle
removable as the number of calibrating Cepheids is increased.

Finally, we observe from table 9 the important fact that the regression
coefficient of $\log P$ for all three samples is
significantly different from the value predicted from equation (21), ie.
-3.33. In fact, if we force the coefficient of $\log P$ to be
-3.33 in our regression fit, we find that the coefficient of
$M_{\mathrm{\ti{mean}}} - < M_{\mathrm{\ti{mean}}}>$
is fitted to be essentially zero. In other words the fit reverts to the
PLC relation at maximum light, with correspondingly increased dispersion.
A similar sensitivity is, in fact,
found with the PLC relation at mean light, for which the theoretically
expected coefficient of $\log P$ is about 4. Forcing the coefficient of
of $\log P$ to be equal to this value results in a
significant change in the $(B-V)_{\mathrm{\ti{mean}}}$
coefficient.

The change in the coefficient of $\log P$ in the PLL'C relation would appear
to be due to the very strong correlation between absolute magnitude at mean
and maximum light (which, as we have discussed, renders the relation nearly
distance degenerate). Thus, while the measured values of $\log P$ and
maximum colour continue to provide useful physical information to constrain
the inferred absolute magnitude at maximum light, the main contribution
to ${{\hat M}_{\mathrm{\ti{max}}}}$ comes from the measured value of
$M_{\mathrm{\ti{mean}}} - < M_{\mathrm{\ti{mean}}}>$ -- and it is the
relationship between these two variables which dominates the form of the fitted
relation. In a sense, therefore, the numerical values of the
coefficients in our PLL'C relation are determined as much by the fact that
we are combining observations of Cepheids at two phase points as by the
underlying physical relationship between period, luminosity and colour.

\section{Conclusions}

Using the period-mean density relation, the Stefan
Boltzmann law and the existence of a linear mass luminosity
law for Cepheids, we have derived a new linear
relation for Cepheids, connecting the maximum and
mean luminosity, the period and effective
temperature. This new equation suggests the possibility of deriving
PLC relations using observations of Cepheids at maximum light, and applying
these relations to estimate the distance of Cepheids.
We have fitted these new relations -- together with
standard PL and PLC relations at mean light -- to a sample of Cepheids
in the LMC. We have adopted a general statistical model which allows for
non-zero correlation between all pairs of the observables, and we have
obtained distance estimators which are `optimal' in the sense of being unbiased
and having minimum variance.

More specifically, our conclusions are the following:-
\begin{enumerate}
\item{Our PL and PLC results at mean light are similar to those given in
MWF and CC. Moreover, we also confirm that the introduction of a
colour term for LMC Cepheids produces a significant reduction
in scatter over a PL relation.}
\item{We have derived a PLC relation for LMC Cepheids at
light maximum. The introduction of the colour term at
maximum light again offers a significant reduction in
scatter compared with the corresponding PL(max) relation. Moreover,
for these data
we find that our PLC(max) relation has around 10 percent less scatter than
a PLC relation at mean light.}
\item{The maximum light, maximum colour, period, semi-amplitude relation
generally
has a comparable dispersion to that obtained for PLC relations at mean or
maximum light, but has a significantly smaller dispersion
when stars with unusual amplitudes for their period are included
in the sample. This result provides evidence
to support the arguments leading to equation (15).}
\item{The maximum light, maximum colour, period and
mean magnitude relation given by equation (39) offers a highly significant
reduction in the dispersion of PLC relations at maximum or mean light. In
converting the relation into a form which is not close to distance degenerate,
we find that
the corresponding distance estimator has a dispersion nearly $40 \%$
smaller than that derived from a PLC relation at maximum light provided
the relation is applied to a sufficiently large ($n = 30$) group of
equidistant Cepheids in, e.g., a distant galaxy: an observational
constraint which does not seem too unreasonable. This relation appears
to derive both from the underlying physical relationship between luminosity,
colour and period, and from the fact that Cepheid observations at two
different phase points are used in its construction. Our results would seem
to offer support for the derivation leading to equation (14),
although further analysis of other data sets would prove very useful in
order to more fully understand this relation.}
\end{enumerate}

\section{Further Work}

We present below some topics which we feel deserve further investigation
based on the results presented in this paper.
\begin{enumerate}
\item{MWF present individual reddening corrections for
many of the stars in table 1. We propose to repeat the
analysis using these reddenings instead of using $E(B-V)=0.1$
for all stars. The use of accurate reddening corrections would be
important in the practical implementation of this
method for distance determinations.
It would also be important to model the
geometry of the LMC, as is carried out in CC.}
\item{CC present data for Cepheids in the
SMC. We have started to further test our results on SMC Cepheids
and other data sets such as those described in
Table (1) of Jacoby et. al (1992).}
\item{We plan to investigate in more detail our 4 variable estimators
using maximum and mean light, period and maximum colour. We would
welcome receiving new Cepheid data sets, such as those available from the
MACHO project and Hubble Telescope observations,
in order to further test and extend the results described here. We also
plan to investigate the construction of PLL'C estimators using
apparent magnitude data at two arbitrary phase points, not
just at mean and maximum light as studied in this paper.}
\item{We have begun a reexamination of calibrating Galactic
Cepheids, as presented in e.g. table 2 of Feast and Walker (1987). Although
Fernie and McGonegal (1983) found that the introduction of a colour term
did not significantly reduce the scatter of the PL relation at mean light,
it would be interesting to apply the results of this paper to
Galactic Cepheids and establish if a significant reduction in dispersion can
be achieved in this case at maximum light.}
\item{Successful completion of (1), (2), (3) and (4) will
enable us to properly calibrate our new relations and compute
a distance modulus to the LMC and SMC.}
\item{Madore and Freedman (1991) show that the dispersion in
Cepheid PL relations at mean light decreases as wavelength
increases. In this work we have used $B-V$ observations.
We plan to extend this by using observations at redder
wavelengths in the hope that this will further reduce
the scatter and in addition serve to reduce the effects of
reddening and metallicity. Thus our method might serve to
further improve the large body of work carried out in the
infrared in the past decade, which has already offered a very sigificant
improvement in Cepheid distance estimators.}
\item{Given that our new estimators using maximum light hold up to
the tests mentioned in (1) and (2) -- and given that they depend upon
the line of sight dispersion in the true distance of the observed Cepheids
being small, it seems appropriate to apply our estimator to
Cepheids in galaxies at cosmological distances.
Prime candidates for such analysis would seem to be IC 4182 (Saha et. al, 1994)
and M100 (Freedman et al, 1994). It should be noted that the former
authors adopted a Galactic metallicity for IC 4182.}
\end{enumerate}

It might be argued that the extra work needed in obtaining light curves
with sufficient phase coverage to define an accurate maximum is
not justified by the relatively small reduction in dispersion when using
maximum light
relations. At least in the case of the LMC, however, the Cepheid data
already obtained from the MACHO (Alcock et. al 1994) project
mean that such observations are already in place.

Much of the above work is well under way and will be reported upon
in subsequent papers.

\section{Acknowledgements}
MAH acknowledges a PPARC personal research fellowship, the use of
the {\sc{starlink}} computer facilities at the
University of Sussex, and the hospitality of the Department of Physics and
Astronomy at the University of Glasgow and the International School for
Advanced Studies in Trieste, where much of this work was carried
out. SMK acknowledges a PPARC grant and several useful discussions with
John Caldwell, Don Fernie and Norman Simon. SMK and MAH thank
Bill Martin for the use of his data and thank the referee for useful
comments.

\section{Appendix: the statistical model}
\appendix{
The distance estimators introduced in Sect. 4 depend upon
various linear combinations of the following six observable
quantities: maximum apparent magnitude, mean apparent magnitude,
maximum colour, mean colour, log period and semi-amplitude. More specifically,
the relations set down in equations (34) - (41) involve using a linear
combination of a subset of
these observables to infer an estimate of the {\em absolute\/} magnitude
at mean or maximum light, which is then combined with the apparent magnitude
to obtain a distance estimator following equation (24) or (25).
In order to compare the variance of these distance estimators, we therefore
require to adopt an appropriate statistical model for the joint distribution of
the six intrinsic physical quantities: maximum absolute magnitude, mean
absolute
magnitude, maximum colour, mean colour, log period and semi-amplitude.
(Of course, since the calibrating Cepheids in the LMC are assumed to be
equidistant, any model for the distribution of mean and maximum absolute
magnitudes will apply equivalently to apparent magnitudes corrected for
absorption).

We adopt a model which is general, realistic and non-restrictive, while
remaining analytically tractable: an obvious candidate is the multivariate
normal distribution. Thus, we assume that the above six physical quantities
are drawn from a distribution denoted by,
\be
{\underline{\bf X}} \sim { N({\underline{\mu}},{\underline{\Sigma}})}
\ee
with probability density function given by,
\be
\biggl({\frac{1}{(2{\pi})^{p/2}}}
{\vert {{\underline{\Sigma}}} \vert}^{1/2}\biggr)
\exp [ {- {\frac{1}{2}}({\bf{\underline X}} - {{\underline{\mu}}})
^{\mathrm{\ti{T}}}
{{\underline{\Sigma}}}^{-1} ({\bf{\underline X}}
- {{\underline{\mu}}}) } ]
\ee
where ${\underline{\bf X}}$ is the vector of variables,
${\underline{\mu}}$ is the vector of their mean values and
${\underline{\Sigma}}$ is the 6$\times$6 covariance matrix describing
their mutual correlation. ${\underline{\Sigma}}$ is assumed to be
positive definite, real, symmetric but is otherwise
arbitrary. In particular, we do {\em not\/} assume ${\underline{\Sigma}}$
to be diagonal: i.e. we allow for all six physical variables to be
correlated with each other -- as clearly may be the case in reality.

Let the elements of
${\underline{\bf X}},{\underline{\mu}}$ and
${\underline{\Sigma}}$ be denoted $(X_1,X_2,...,X_6)$,
$({\mu}_1,{\mu}_2,...,{\mu}_6)$ and ${\sigma}_{ij}$ respectively.
The statistical problem with which we are dealing concerns the optimal
estimation, or {\em prediction\/}, of $X_j$  for some $j$ (denoting the mean or
maximum absolute magnitude) from a linear combination of the measured values
of some or all of the other variables,
$X_i$. This is an example of a {\em linear prediction\/} problem - a topic
which
is treated extensively in the general statistics literature
(c.f. Kendall and Stuart 1963; Graybill, 1976)

Following the standard notation, we define the best prediction function
$g_1(X_2,...,X_6)$ of, say, $X_1$
based upon the measured values of $X_2,....,X_6$ to be the function such that,
\be
E \{ {[X_1 - g_1(X_2,...,X_6)]}^2 \} \leq E \{ {[X_1 - g(X_2,....,X_6)]}^2 \}
\ee
for all other such functions, $g(X_2,...X_6)$, of $(X_2,...,X_6)$
(c.f. eqn (12.2.2) of Graybill, 1976). That is, the best prediction function is
`best' in the
sense of minimum variance and unbiased - i.e. minimum risk, as we define in
Sect. 4.

It can be shown (c.f. Graybill 1976; Kendall and Stuart 1963) that when
${\underline{\bf X}}$ has a multivariate normal distribution, the
best {\em linear} prediction function of $X_1$ based upon the measured values
of
$X_2,....,X_6$ is
\be
E(X_1 | X_2=x_2,....X_6=x_6) = {\beta}_1 + {\sum^6_{k=2}} {\beta}_kx_k
\ee
where $x_j$ denotes the measured value of the variable $X_j$, and the
regression coefficient, ${\beta}_j$, is defined by,
\be
{\beta}_j =
{\frac{{\mathrm{cov}}[(X_1,X_j | X_2,..,X_{j-1},X_{j+1},..,X_6)]}
{{\mathrm{var}}[(X_j | X_2,...,X_{j-1},X_{j+1},...X_6)]}}
\ee

In other words, the best linear prediction function of $X_1$ is given
simply by the
linear regression of $X_1$ on $X_2,....X_6$. Obviously the equivalent result
holds for the best linear predictor of the other $X_j$. Moreover, the result
also holds in the case where the covariance matrix, ${\underline{\Sigma}}$,
is not known
a priori but must be estimated from observations - as is obviously the case
in the current study. In this event, the covariance matrix - and the
regression coefficients in equation (A.5) calculated from it - are
simply replaced by their sample estimates.

Our procedure for finding the optimum set of
predicition variables for
$M_{\mathrm{\ti{max}}}$ or
$M_{\mathrm{\ti{mean}}}$ is based upon
the forward selection procedure, as described in e.g. Chapter 12 of Graybill
(1976). Again, we outline this procedure where one is predicting $X_1$
from measurements of the other variables, with the obvious equivalent
procedure for predicting some other $X_j$.

First, we determine the sample multiple correlation coefficient,
${\rho}_{1(k)}$, of
$X_1$ with $X_k$ for $k=2,...,6$, and identify the largest of
these in modulus. Suppose, without loss of generality, that this is
${\rho}_{1(2)}$.
Then $X_2$ is the best single predictor of $X_1$. We then compute
all multiple correlation coefficients with $X_1$ of {\em pairs\/} of variables
which include the best single predictor, $X_2$: i.e. ${\rho_{1(2,3)}}$,
${\rho_{1(2,4)}},...,{\rho_{1(2,6)}}$.
Again, we select the largest - say ${\rho_{1(2,s)}}$.
It then follows that $X_2$ and $X_s$ are the
best two predictors of $X_1$ which include the best single
predictor $X_2$.

We can continue this procedure until all the remaining
variables have been included in the prediction function or, more usefully,
until the addition of a new variable
does not appreciably improve the estimator of $X_1$. We can test the
significance of adding a given variable -- and hence quantify what we mean
by `appreciably improve' the fit -- in two different ways. First,
we can test the null hypothesis that the regression coefficient of a
given variable, $X_j$, is equal to zero. i.e.,
\be
H_0 : {\beta}_j = 0
\ee

If the null hypothesis is strongly rejected, then one should include
$X_j$ in the best linear prediction function. If the null hypothesis
is accepted, then $X_j$ has little effect on the prediction of $X_1$ and
could be considered superfluous. It can be shown (c.f. Graybill, 1976)
that under the null hypothesis
the transformed variable
$t_j = {\hat{{\beta}_j}}/{\hat{\sigma}}({\hat{{\beta}_j}})$
has a t distribution with $n - \alpha$ degrees of freedom. Here
${\hat{\sigma}}$
denotes the standard error of the estimated regression coefficient
and
$n$ and ${\alpha}$ are the number of data points in the linear regression fit
and the number of independent variables in the fit (i.e. the number of
prediction variables) respectively. Testing the null hypothesis that
${\beta}_j = 0$
is therefore equivalent to the test,
\be
H_0 : t_j = 0
\ee
the results of which for each of our estimators are described in detail in
Sect. 5.

We can also test the significance of adding a given variable in terms of the
computed sample multiple correlation coefficients of the different relations.
Suppose we have applied the forward selection procedure
described above to identify the best linear prediction function of $X_1$
as depending on the variables $X_2,...,X_q$, and we wish to test whether the
new variable, $X_{q+1}$ significantly improves the estimation of $X_1$.
If $X_{q+1}$ has {\em no\/} effect on the prediction then the multiple
correlation coefficient of $X_1$ with $X_2,...,X_{q}$ will be identically
equal to that of $X_1$ with $X_2,...,X_{q+1}$, i.e.,
\be
{\rho}^2_{1(2,...q+1)} = {\rho}^2_{1(2,...q)}
\ee

One can then use
the sampled values of the multiple correlation coefficients to construct a
suitable hypothesis test based upon this property, i.e. to test the
null hypothesis,
\be
H_0 : {\hat{\rho}}^2_{1(2,...q+1)} = {\hat{\rho}}^2_{1(2,...q)}
\ee
where the caret denotes that the multiple correlation coefficients are
sample values estimated from the calibration data.

In fact, the test in equation (A.9) is
equivalent to testing the null hypothesis,
\be
H_0 : {\rho}^2_{1,q+1 \vert (2,...q)} = 0
\ee
where ${\rho_{1,q+1 \vert (2,...q)}}$ is the partial sample
multiple correlation coefficient of $X_1$ and $X_{q+1}$ given
$X_2,...X_q$.

When $H_0$ is true, equation (A.10) is
distributed as a central F random variable with 1 and $n-q-2$
degrees of freedom. It is this hypothesis test which we have applied
to each of the relations introduced in Sect. 4.

Since the assumption of a multivariate normal distribution is required in
most of the above, we have devoted Sect. 6 to the task of testing carefully
how well our LMC data satisfy this normality assumption.}
\clearpage
\newpage

\begin{center}
{\bf{\underline{Table Captions}}}
\end{center}
\vspace{1cm}

\noindent Table 1 : LMC stars
\vspace{4mm}

\noindent Table 2 : Excluded LMC stars
\vspace{4mm}

\noindent Table 3 : Uunusual LMC stars
\vspace{4mm}

\noindent Table 4 : Results of KS test applied to LMC data
\vspace{4mm}

\noindent Table 5 : Skewness and kurtosis results
\vspace{4mm}

\noindent Table 6 : Shapiro and Wilks normality test results
\vspace{4mm}

\noindent Table 7 : Fitted PL relations
\vspace{4mm}

\noindent Table 8 : Fitted PLC and PLA relations
\vspace{4mm}

\noindent Table 9 : Fitted PLL'C and PLAC relations
\clearpage
\newpage
\footnotesize

\begin{center}
{\bf{\underline{TABLE 1}}}
\vspace{4mm}

{\underline{LMC STARS}}
\vspace{3mm}

\begin{tabular}[t]{ll}
Star&Period (days)\\ \hline
HV2353&3.1080\\
HV12765&3.4290\\
HV12700&8.1530\\
HV12823&8.3020\\
HV2854&8.6350\\
HV2733&8.7220\\
HV12816&9.1140\\
HV971&9.2970\\
HV2301&9.4990\\
HV6105&10.4400\\
HV2864&10.9840\\
HV874&12.6820\\
HV2260&12.9360\\
HV2527&12.9480\\
HV997&13.1470\\
HV2579&13.4310\\
HV2352&13.6260\\
HV955&13.7320\\
HV2324&14.4660\\
HV2549&16.1970\\
HV2580&16.9450\\
HV2836&17.5260\\
HV1005&18.7100\\
HV2793&19.1840\\
HV1013&24.1264\\
HV12815&26.1690\\
HV1023&26.5880\\
HV1002&30.4700\\
HV899&31.0270\\
HV2294&36.5270\\
HV2294&36.5270\\
HV879&36.7820\\
HV2338&42.1669\\
HV877&45.1853\\
HV2369&48.3190\\
HV2827&78.8582\\
HV5497&98.7802\\
HV2883&109.000\\
HV2447&119.4400\\
HV883&134.000\\
\end{tabular}
\end{center}
\clearpage
\newpage

\begin{center}
{\bf{\underline{TABLE 2}}}
\vspace{4mm}

{\underline{EXCLUDED LMC STARS}}
\vspace{3mm}

\begin{tabular}[t]{ll}
Star&Period (days)\\ \hline
HV5655&14.2110\\
HV2262&15.8460\\
HV909&37.5700\\
HV2257&42.1669\\
HV900&47.5330\\
HV953&47.890\\
\end{tabular}
\end{center}
\vspace{1cm}

\begin{center}
{\bf{\underline{TABLE 3}}}
\vspace{4mm}

{\underline{UNUSUAL LMC STARS}}
\vspace{3mm}

\begin{tabular}[t]{ll}
Star&Period (days)\\ \hline
HV877&45.1853\\
HV2827&78.8582\\
HV5497&98.7802\\
HV2883&109.000\\
HV2447&119.440\\
HV883&134.000\\
\end{tabular}
\end{center}
\clearpage
\newpage

\begin{center}
{\bf{\underline{TABLE 4}}}
\vspace{4mm}

{\underline{RESULTS OF KS TEST APPLIED TO LMC DATA}}
\end{center}
\vspace{3mm}

\begin{tabular}[t]{lrcc}
& Variable & $D_{\mathop{\rm{obs}}}$ &
Prob($D_{\mathop{\rm{n}}} > D_{\mathop{\rm{obs}}}$)\\ \hline
 31 stars & $V_{\mathop{\rm{max}}}$      & 0.099 & 0.905 \\
          & $V_{\mathop{\rm{mean}}}$     & 0.103 & 0.880 \\
          & $(B-V)_{\mathop{\rm{max}}}$  & 0.096 & 0.927 \\
          & $(B-V)_{\mathop{\rm{mean}}}$ & 0.099 & 0.907 \\
          & $\log P$                     & 0.125 & 0.688 \\
          & semi-amplitude               & 0.144 & 0.504 \\
          & & & \\
          & & & \\
 35 stars & $V_{\mathop{\rm{max}}}$      & 0.100 & 0.858 \\
          & $V_{\mathop{\rm{mean}}}$     & 0.120 & 0.663 \\
          & $(B-V)_{\mathop{\rm{max}}}$  & 0.120 & 0.663 \\
          & $(B-V)_{\mathop{\rm{mean}}}$ & 0.098 & 0.875 \\
          & $\log P$                     & 0.121 & 0.651 \\
          & semi-amplitude               & 0.151 & 0.372 \\
          & & & \\
          & & & \\
 39 stars & $V_{\mathop{\rm{max}}}$      & 0.111 & 0.697 \\
          & $V_{\mathop{\rm{mean}}}$     & 0.152 & 0.304 \\
          & $(B-V)_{\mathop{\rm{max}}}$  & 0.172 & 0.179 \\
          & $(B-V)_{\mathop{\rm{mean}}}$ & 0.100 & 0.798 \\
          & $\log P$                     & 0.124 & 0.562 \\
          & semi-amplitude               & 0.150 & 0.315 \\
\end{tabular}
\clearpage
\newpage

\begin{center}
{\bf{\underline{TABLE 5}}}
\vspace{4mm}

{\underline{SKEWNESS AND KURTOSIS RESULTS}}
\end{center}
\vspace{3mm}

\begin{tabular}[t]{lrcc}
 & Variable & Skewness (no. of $\sigma$) &
Kurtosis (no. of $\sigma$)\\ \hline
 31 stars & $V_{\mathop{\rm{max}}}$      & 0.56 & 0.04 \\
          & $V_{\mathop{\rm{mean}}}$     & 0.72 & 0.01 \\
          & $(B-V)_{\mathop{\rm{max}}}$  & 0.57 & 0.38 \\
          & $(B-V)_{\mathop{\rm{mean}}}$ & 0.25 & 0.42 \\
          & $\log P$                     & 0.72 & 0.29 \\
          & semi-amplitude               & 0.52 & 0.58 \\
          & & & \\
          & & & \\
 35 stars & $V_{\mathop{\rm{max}}}$      & 0.77 & 0.18 \\
          & $V_{\mathop{\rm{mean}}}$     & 0.98 & 0.11 \\
          & $(B-V)_{\mathop{\rm{max}}}$  & 1.51 & 0.63 \\
          & $(B-V)_{\mathop{\rm{mean}}}$ & 0.32 & 0.43 \\
          & $\log P$                     & 0.46 & 0.51 \\
          & semi-amplitude               & 0.61 & 0.60 \\
          & & & \\
          & & & \\
 39 stars & $V_{\mathop{\rm{max}}}$      & 0.60 & 0.56 \\
          & $V_{\mathop{\rm{mean}}}$     & 1.02 & 0.40 \\
          & $(B-V)_{\mathop{\rm{max}}}$  & 1.68 & 0.10 \\
          & $(B-V)_{\mathop{\rm{mean}}}$ & 0.15 & 0.65 \\
          & $\log P$                     & 0.66 & 0.09 \\
          & semi-amplitude               & 0.46 & 0.74 \\
\end{tabular}
\clearpage
\newpage

\begin{center}
{\bf{\underline{TABLE 6}}}
\vspace{4mm}

{\underline{SHAPIRO AND WILKS NORMALITY TEST RESULTS}}
\end{center}
\vspace{3mm}

\begin{tabular}[t]{lrcc}
 & Variable & $W_{\mathop{\rm{obs}}}$ & Significance\\ \hline
 31 stars & $V_{\mathop{\rm{max}}}$      & 0.977 & 0.7546 \\
          & $V_{\mathop{\rm{mean}}}$     & 0.961 & 0.3630 \\
          & $(B-V)_{\mathop{\rm{max}}}$  & 0.954 & 0.2432 \\
          & $(B-V)_{\mathop{\rm{mean}}}$ & 0.959 & 0.3117 \\
          & $\log P$                     & 0.944 & 0.1281 \\
          & semi-amplitude               & 0.927 & 0.0431 \\
          & & & \\
          & & & \\
 35 stars & $V_{\mathop{\rm{max}}}$      & 0.960 & 0.3010 \\
          & $V_{\mathop{\rm{mean}}}$     & 0.948 & 0.1276 \\
          & $(B-V)_{\mathop{\rm{max}}}$  & 0.928 & 0.0301 \\
          & $(B-V)_{\mathop{\rm{mean}}}$ & 0.972 & 0.5873 \\
          & $\log P$                     & 0.965 & 0.3850 \\
          & semi-amplitude               & 0.927 & 0.0283 \\
          & & & \\
          & & & \\
 39 stars & $V_{\mathop{\rm{max}}}$      & 0.940 & 0.0508 \\
          & $V_{\mathop{\rm{mean}}}$     & 0.919 & 0.0090 \\
          & $(B-V)_{\mathop{\rm{max}}}$  & 0.865 & 0.0001 \\
          & $(B-V)_{\mathop{\rm{mean}}}$ & 0.953 & 0.1469 \\
          & $\log P$                     & 0.948 & 0.0972 \\
          & semi-amplitude               & 0.928 & 0.0197 \\
\end{tabular}
\clearpage
\newpage

\begin{center}
{\bf{TABLE 7}}
\end{center}
\vspace{1mm}

\begin{center}
{FITTED PL RELATIONS}
\end{center}

\noindent
{\footnotesize{\underline{31 stars}}}
\vspace{2mm}

{$\hat M = a + b \log P$}
\vspace{2mm}

\begin{tabular}[t]{crrcrrr}
& \hspace{6mm} ${\hat{\beta}}$ \hspace{2mm}
& \hspace{7mm} ${\sigma}_{\hat{\beta}}$ \hspace{1mm}
& \hspace{6mm} {\footnotesize{{$\log P(\beta = 0)$}}} \hspace{2mm}
& \hspace{7mm} ${\sigma}_{{\hat M}_*}$ \hspace{1mm}
& \hspace{4mm} $\Delta$ \hspace{4mm}\\ \hline
a & -1.90 & 0.27 & & 0.324 & 14.9\\
b & -2.05 & 0.23 & -9.36 & & \\ \hline
\end{tabular}
\vspace{5mm}

{$\hat M_{\mathop{\rm{max}}} = a + b \log P$}
\vspace{2mm}

\begin{tabular}[t]{crrcrrr}
& \hspace{6mm} ${\hat{\beta}}$ \hspace{2mm}
& \hspace{7mm} ${\sigma}_{\hat{\beta}}$ \hspace{1mm}
& \hspace{6mm} {\footnotesize{{$\log P(\beta = 0)$}}} \hspace{2mm}
& \hspace{7mm} ${\sigma}_{{\hat M}_*}$ \hspace{1mm}
& \hspace{4mm} $\Delta$ \hspace{4mm}\\ \hline
a & -1.88 & 0.26 & & 0.316 & 14.5\\
b & -2.48 & 0.22 & $<$ -10 & & \\ \hline
\end{tabular}
\vspace{8mm}

\noindent
{\footnotesize{\underline{35 stars}}}
\vspace{2mm}

{$\hat M = a + b \log P$}
\vspace{2mm}

\begin{tabular}[t]{crrcrrr}
& \hspace{6mm} ${\hat{\beta}}$ \hspace{2mm}
& \hspace{7mm} ${\sigma}_{\hat{\beta}}$ \hspace{1mm}
& \hspace{6mm} {\footnotesize{{$\log P(\beta = 0)$}}} \hspace{2mm}
& \hspace{7mm} ${\sigma}_{{\hat M}_*}$ \hspace{1mm}
& \hspace{4mm} $\Delta$ \hspace{4mm}\\ \hline
a & -1.80 & 0.21 & & 0.323 & 14.9\\
b & -2.16 & 0.17 & $<$ -10 & & \\ \hline
\end{tabular}
\vspace{5mm}

{$\hat M_{\mathop{\rm{max}}} = a + b \log P$}
\vspace{2mm}

\begin{tabular}[t]{crrcrrr}
& \hspace{6mm} ${\hat{\beta}}$ \hspace{2mm}
& \hspace{7mm} ${\sigma}_{\hat{\beta}}$ \hspace{1mm}
& \hspace{6mm} {\footnotesize{{$\log P(\beta = 0)$}}} \hspace{2mm}
& \hspace{7mm} ${\sigma}_{{\hat M}_*}$ \hspace{1mm}
& \hspace{4mm} $\Delta$ \hspace{4mm}\\ \hline
a & -1.93 & 0.22 & & 0.326 & 15.0\\
b & -2.44 & 0.17 & $<$ -10 & & \\ \hline
\end{tabular}
\vspace{8mm}

\noindent
{\footnotesize{\underline{39 stars}}}
\vspace{2mm}

{$\hat M = a + b \log P$}
\vspace{2mm}

\begin{tabular}[t]{crrcrrr}
& \hspace{6mm} ${\hat{\beta}}$ \hspace{2mm}
& \hspace{7mm} ${\sigma}_{\hat{\beta}}$ \hspace{1mm}
& \hspace{6mm} {\footnotesize{{$\log P(\beta = 0)$}}} \hspace{2mm}
& \hspace{7mm} ${\sigma}_{{\hat M}_*}$ \hspace{1mm}
& \hspace{4mm} $\Delta$ \hspace{4mm}\\ \hline
a & -1.64 & 0.18 & & 0.317 & 14.6\\
b & -2.30 & 0.13 & $<$ -10 & & \\ \hline
\end{tabular}
\vspace{5mm}

{$\hat M_{\mathop{\rm{max}}} = a + b \log P$}
\vspace{2mm}

\begin{tabular}[t]{crrcrrr}
& \hspace{6mm} ${\hat{\beta}}$ \hspace{2mm}
& \hspace{7mm} ${\sigma}_{\hat{\beta}}$ \hspace{1mm}
& \hspace{6mm} {\footnotesize{{$\log P(\beta = 0)$}}} \hspace{2mm}
& \hspace{7mm} ${\sigma}_{{\hat M}_*}$ \hspace{1mm}
& \hspace{4mm} $\Delta$ \hspace{4mm}\\ \hline
a & -2.04 & 0.18 & & 0.316 & 14.5\\
b & -2.34 & 0.13 & $<$ -10 & & \\ \hline
\end{tabular}
\clearpage
\newpage

\begin{center}
{\bf{TABLE 8}}
\end{center}
\vspace{1mm}

\begin{center}
{FITTED PLC AND PLA RELATIONS}
\end{center}

\noindent
{\footnotesize{\underline{31 stars}}}
\vspace{2mm}

{$\hat M = a + b \log P + c (B-V)$}
\vspace{2mm}

\begin{tabular}[t]{crrcrr}
& \hspace{6mm} ${\hat{\beta}}$ \hspace{2mm}
& \hspace{7mm} ${\sigma}_{\hat{\beta}}$ \hspace{1mm}
& \hspace{6mm} {\footnotesize{{$\log P(\beta = 0)$}}} \hspace{2mm}
& \hspace{7mm} ${\sigma}_{{\hat M}_*}$ \hspace{1mm}
& \hspace{4mm} $\Delta$ \hspace{4mm}\\ \hline
a & -2.11 & 0.23 & & 0.265 & 12.2\\
b & -3.03 & 0.31 & $<$ -10 & & \\
c &  1.75 & 0.44 & -3.61 & & \\ \hline
\end{tabular}
\vspace{0.8mm}

partial SMCC = 0.597 \hspace{1.5cm}
P($H_0$) = $4.9 \times 10^{-4}$
\vspace{5mm}

{$\hat M_{\mathop{\rm{max}}} = a + b \log P + c (B-V)_{\mathop{\rm{max}}}$}
\vspace{2mm}

\begin{tabular}[t]{crrcrr}
& \hspace{6mm} ${\hat{\beta}}$ \hspace{2mm}
& \hspace{7mm} ${\sigma}_{\hat{\beta}}$ \hspace{1mm}
& \hspace{6mm} {\footnotesize{{$\log P(\beta = 0)$}}} \hspace{2mm}
& \hspace{7mm} ${\sigma}_{{\hat M}_*}$ \hspace{1mm}
& \hspace{4mm} $\Delta$ \hspace{4mm}\\ \hline
a & -2.39 & 0.22 & & 0.235 & 10.8\\
b & -2.87 & 0.18 & $<$ -10 & & \\
c &  2.05 & 0.41 & -4.80 & & \\ \hline
\end{tabular}
\vspace{0.8mm}

partial SMCC = 0.683 \hspace{1.5cm}
P($H_0$) = $3.2 \times 10^{-5}$
\vspace{8mm}

\noindent
{\footnotesize{\underline{35 stars}}}
\vspace{2mm}

{$\hat M = a + b \log P + c (B-V)$}
\vspace{2mm}

\begin{tabular}[t]{crrcrr}
& \hspace{6mm} ${\hat{\beta}}$ \hspace{2mm}
& \hspace{7mm} ${\sigma}_{\hat{\beta}}$ \hspace{1mm}
& \hspace{6mm} {\footnotesize{{$\log P(\beta = 0)$}}} \hspace{2mm}
& \hspace{7mm} ${\sigma}_{{\hat M}_*}$ \hspace{1mm}
& \hspace{4mm} $\Delta$ \hspace{4mm}\\ \hline
a & -2.09 & 0.18 & & 0.257 & 11.8\\
b & -3.11 & 0.25 & $<$ -10 & & \\
c &  1.83 & 0.41 & -4.37 & & \\ \hline
\end{tabular}
\vspace{0.8mm}

partial SMCC = 0.622 \hspace{1.5cm}
P($H_0$) = $8.6 \times 10^{-5}$
\vspace{5mm}

{$\hat M_{\mathop{\rm{max}}} = a + b \log P + c (B-V)_{\mathop{\rm{max}}}$}
\vspace{2mm}

\begin{tabular}[t]{crrcrr}
& \hspace{6mm} ${\hat{\beta}}$ \hspace{2mm}
& \hspace{7mm} ${\sigma}_{\hat{\beta}}$ \hspace{1mm}
& \hspace{6mm} {\footnotesize{{$\log P(\beta = 0)$}}} \hspace{2mm}
& \hspace{7mm} ${\sigma}_{{\hat M}_*}$ \hspace{1mm}
& \hspace{4mm} $\Delta$ \hspace{4mm}\\ \hline
a & -2.32 & 0.16 & & 0.227 & 10.5\\
b & -2.92 & 0.14 & $<$ -10 & & \\
c &  2.00 & 0.33 & -6.25 & & \\ \hline
\end{tabular}
\vspace{0.8mm}

partial SMCC =  0.727 \hspace{1.5cm}
P($H_0$) = $1.1 \times 10^{-6}$
\vspace{8mm}

\clearpage
\newpage

\begin{center}
{\bf{TABLE 8 continued}}
\end{center}
\vspace{1mm}

\noindent
{\footnotesize{\underline{39 stars}}}
\vspace{2mm}

{$\hat M = a + b \log P + c (B-V)$}
\vspace{2mm}

\begin{tabular}[t]{crrcrr}
& \hspace{6mm} ${\hat{\beta}}$ \hspace{2mm}
& \hspace{7mm} ${\sigma}_{\hat{\beta}}$ \hspace{1mm}
& \hspace{6mm} {\footnotesize{{$\log P(\beta = 0)$}}} \hspace{2mm}
& \hspace{7mm} ${\sigma}_{{\hat M}_*}$ \hspace{1mm}
& \hspace{4mm} $\Delta$ \hspace{4mm}\\ \hline
a & -2.01 & 0.16 & & 0.251 & 11.6\\
b & -3.22 & 0.22 & $<$ -10 & & \\
c &  1.88 & 0.39 & -4.83 & & \\ \hline
\end{tabular}
\vspace{0.8mm}

partial SMCC = 0.623 \hspace{1.5cm}
P($H_0$) = $3.0 \times 10^{-5}$
\vspace{5mm}

{$\hat M_{\mathop{\rm{max}}} = a + b \log P + c (B-V)_{\mathop{\rm{max}}}$}
\vspace{2mm}

\begin{tabular}[t]{crrcrr}
& \hspace{6mm} ${\hat{\beta}}$ \hspace{2mm}
& \hspace{7mm} ${\sigma}_{\hat{\beta}}$ \hspace{1mm}
& \hspace{6mm} {\footnotesize{{$\log P(\beta = 0)$}}} \hspace{2mm}
& \hspace{7mm} ${\sigma}_{{\hat M}_*}$ \hspace{1mm}
& \hspace{4mm} $\Delta$ \hspace{4mm}\\ \hline
a & -2.11 & 0.13 & & 0.234 & 10.8\\
b & -2.96 & 0.15 & $<$ -10 & & \\
c & 1.64 & 0.29 & -5.92 & & \\ \hline
\end{tabular}
\vspace{0.8mm}

partial SMCC =  0.682 \hspace{1.5cm}
P($H_0$) = $2.4 \times 10^{-6}$
\vspace{5mm}

\noindent
{\footnotesize{\underline{31 stars}}}
\vspace{2mm}

{$\hat M = a + b \log P + c (M_{\mathop{\rm{mean}}} - M_{\mathop{\rm{max}}})$}
\vspace{2mm}

\begin{tabular}[t]{crrcrr}
& \hspace{6mm} ${\hat{\beta}}$ \hspace{2mm}
& \hspace{7mm} ${\sigma}_{\hat{\beta}}$ \hspace{1mm}
& \hspace{6mm} {\footnotesize{{$\log P(\beta = 0)$}}} \hspace{2mm}
& \hspace{7mm} ${\sigma}_{{\hat M}_*}$ \hspace{1mm}
& \hspace{4mm} $\Delta$ \hspace{4mm}\\ \hline
a & -1.88 & 0.27 & & 0.321 & 14.8\\
b & -2.40 & 0.36 & -6.77 & & \\
c & -0.19 & 0.65 & -0.41 & & \\ \hline
\end{tabular}
\vspace{0.8mm}

partial SMCC = 0.230 \hspace{1.5cm}
P($H_0$) = $0.222$
\vspace{5mm}

{$\hat M_{\mathop{\rm{max}}} = a + b \log P + c (M_{\mathop{\rm{mean}}} -
M_{\mathop{\rm{max}}})$}
\vspace{2mm}

\begin{tabular}[t]{crrcrr}
& \hspace{6mm} ${\hat{\beta}}$ \hspace{2mm}
& \hspace{7mm} ${\sigma}_{\hat{\beta}}$ \hspace{1mm}
& \hspace{6mm} {\footnotesize{{$\log P(\beta = 0)$}}} \hspace{2mm}
& \hspace{7mm} ${\sigma}_{{\hat M}_*}$ \hspace{1mm}
& \hspace{4mm} $\Delta$ \hspace{4mm}\\ \hline
a & -1.88 & 0.27 & & 0.321 & 14.8\\
b & -2.40 & 0.36 & -6.77 & & \\
c &  0.81 & 0.65 & -0.96 & & \\ \hline
\end{tabular}
\vspace{0.8mm}

partial SMCC = -0.054 \hspace{1.5cm}
P($H_0$) = $0.776$
\vspace{8mm}

\clearpage
\newpage

\begin{center}
{\bf{TABLE 8 continued}}
\end{center}
\vspace{1mm}

\noindent
{\footnotesize{\underline{35 stars}}}
\vspace{2mm}

{$\hat M = a + b \log P + c (M_{\mathop{\rm{mean}}} - M_{\mathop{\rm{max}}}) $}
\vspace{2mm}

\begin{tabular}[t]{crrcrr}
& \hspace{6mm} ${\hat{\beta}}$ \hspace{2mm}
& \hspace{7mm} ${\sigma}_{\hat{\beta}}$ \hspace{1mm}
& \hspace{6mm} {\footnotesize{{$\log P(\beta = 0)$}}} \hspace{2mm}
& \hspace{7mm} ${\sigma}_{{\hat M}_*}$ \hspace{1mm}
& \hspace{4mm} $\Delta$ \hspace{4mm}\\ \hline
a & -1.85 & 0.23 & & 0.325 & 15.0\\
b & -2.28 & 0.22 & $<$ -10 & & \\
c &  0.44 & 0.51 & -0.70 & & \\ \hline
\end{tabular}
\vspace{0.8mm}

partial SMCC = 0.150 \hspace{1.5cm}
P($H_0$) = $0.4 \times 10^{-5}$
\vspace{5mm}

{$\hat M_{\mathop{\rm{max}}} = a + b \log P + c (M_{\mathop{\rm{mean}}} -
M_{\mathop{\rm{max}}})$}
\vspace{2mm}

\begin{tabular}[t]{crrcrr}
& \hspace{6mm} ${\hat{\beta}}$ \hspace{2mm}
& \hspace{7mm} ${\sigma}_{\hat{\beta}}$ \hspace{1mm}
& \hspace{6mm} {\footnotesize{{$\log P(\beta = 0)$}}} \hspace{2mm}
& \hspace{7mm} ${\sigma}_{{\hat M}_*}$ \hspace{1mm}
& \hspace{4mm} $\Delta$ \hspace{4mm}\\ \hline
a & -1.85 & 0.23 & & 0.325 & 15.0\\
b & -2.28 & 0.22 & $<$ -10 & & \\
c & -0.56 & 0.51 & -0.85 & & \\ \hline
\end{tabular}
\vspace{0.8mm}

partial SMCC = -0.190 \hspace{1.5cm}
P($H_0$) = $0.283$
\vspace{8mm}

\noindent
{\footnotesize{\underline{39 stars}}}
\vspace{2mm}

{$\hat M = a + b \log P + c (M_{\mathop{\rm{mean}}} - M_{\mathop{\rm{max}}})$}
\vspace{2mm}

\begin{tabular}[t]{crrcrr}
& \hspace{6mm} ${\hat{\beta}}$ \hspace{2mm}
& \hspace{7mm} ${\sigma}_{\hat{\beta}}$ \hspace{1mm}
& \hspace{6mm} {\footnotesize{{$\log P(\beta = 0)$}}} \hspace{2mm}
& \hspace{7mm} ${\sigma}_{{\hat M}_*}$ \hspace{1mm}
& \hspace{4mm} $\Delta$ \hspace{4mm}\\ \hline
a & -1.84 & 0.21 & & 0.310 & 14.3\\
b & -2.32 & 0.13 & $<$ -10 & & \\
c &  0.51 & 0.32 & -1.21 & & \\ \hline
\end{tabular}
\vspace{0.8mm}

partial SMCC = 0.254 \hspace{1.5cm}
P($H_0$) = $0.124$
\vspace{5mm}

{$\hat M_{\mathop{\rm{max}}} = a + b \log P + c (M_{\mathop{\rm{max}}} -
M_{\mathop{\rm{mean}}})$}
\vspace{2mm}

\begin{tabular}[t]{crrcrr}
& \hspace{6mm} ${\hat{\beta}}$ \hspace{2mm}
& \hspace{7mm} ${\sigma}_{\hat{\beta}}$ \hspace{1mm}
& \hspace{6mm} {\footnotesize{{$\log P(\beta = 0)$}}} \hspace{2mm}
& \hspace{7mm} ${\sigma}_{{\hat M}_*}$ \hspace{1mm}
& \hspace{4mm} $\Delta$ \hspace{4mm}\\ \hline
a & -1.84 & 0.21 & & 0.310 & 14.3\\
b & -2.32 & 0.13 & $<$ -10 & & \\
c & -0.49 & 0.32 & -1.52 & & \\ \hline
\end{tabular}
\vspace{0.8mm}

partial SMCC =  -0.245 \hspace{1.5cm}
P($H_0$) = $0.138$
\clearpage
\newpage

\begin{center}
{\bf{TABLE 9}}
\end{center}
\vspace{1mm}

\begin{center}
{FITTED PLL'C AND PLAC RELATIONS}
\end{center}

\noindent
{\footnotesize{\underline{31 stars}}}
\vspace{2mm}

{$\hat M_{\mathop{\rm{max}}} = a + b \log P + c (B-V)
+ d (M_{\mathop{\rm{mean}}} - < M_{\mathop{\rm{mean}}}>)$}
\vspace{2mm}

\begin{tabular}[t]{crrcrr}
& \hspace{6mm} ${\hat{\beta}}$ \hspace{2mm}
& \hspace{7mm} ${\sigma}_{\hat{\beta}}$ \hspace{1mm}
& \hspace{6mm} {\footnotesize{{$\log P(\beta = 0)$}}} \hspace{2mm}
& \hspace{7mm} ${\sigma}_{{\hat M}_*}$ \hspace{1mm}
& \hspace{4mm} \hspace{4mm}\\ \hline
a & -3.95 & 0.89 & & 0.078  \\
b & -0.89 & 0.14 & -6.19 & & \\
c &  0.57 & 0.17 & -2.94 & & \\
d &  0.83 & 0.05 &$< -10$ & & \\ \hline
\end{tabular}
\vspace{0.8mm}

partial SMCC = 0.946 \hspace{1.5cm}
P($H_0$) = $0.9 \times 10^{-15}$
\vspace{5mm}

{$\hat M_{\mathop{\rm{max}}} = a + b \log P + c (B-V)_{\mathop{\rm{max}}}
+ d (M_{\mathop{\rm{mean}}} - M_{\mathop{\rm{max}}})$}
\vspace{2mm}

\begin{tabular}[t]{crrcrr}
& \hspace{6mm} ${\hat{\beta}}$ \hspace{2mm}
& \hspace{7mm} ${\sigma}_{\hat{\beta}}$ \hspace{1mm}
& \hspace{6mm} {\footnotesize{{$\log P(\beta = 0)$}}} \hspace{2mm}
& \hspace{7mm} ${\sigma}_{{\hat M}_*}$ \hspace{1mm}
& \hspace{4mm} $\Delta$ \hspace{4mm}\\ \hline
a & -2.41 & 0.22 & & 0.234 & 10.7\\
b & -3.14 & 0.30 & $<$ -10 & & \\
c &  2.19 & 0.43 & -4.92 & & \\
d &  0.56 & 0.50 & -0.87 & & \\ \hline
\end{tabular}
\vspace{0.8mm}

partial SMCC = 0.213 \hspace{1.5cm}
P($H_0$) = $0.267$
\vspace{8mm}

\noindent
{\footnotesize{\underline{35 stars}}}
\vspace{2mm}

{$\hat M_{\mathop{\rm{max}}} = a + b \log P + c (B-V)
+ d (M_{\mathop{\rm{mean}}} - < M_{\mathop{\rm{mean}}}>)$}
\vspace{2mm}

\begin{tabular}[t]{crrcrr}
& \hspace{6mm} ${\hat{\beta}}$ \hspace{2mm}
& \hspace{7mm} ${\sigma}_{\hat{\beta}}$ \hspace{1mm}
& \hspace{6mm} {\footnotesize{{$\log P(\beta = 0)$}}} \hspace{2mm}
& \hspace{7mm} ${\sigma}_{{\hat M}_*}$ \hspace{1mm}
& \hspace{4mm} \hspace{4mm}\\ \hline
a & -4.13 & 0.90 & & 0.083 & \\
b & -0.92 & 0.15 & -6.40 & & \\
c &  0.76 & 0.15 & -5.03 & & \\
d &  0.79 & 0.05 & $< -10$& & \\ \hline
\end{tabular}
\vspace{0.8mm}

partial SMCC = 0.932 \hspace{1.5cm}
P($H_0$) = $3.1 \times 10^{-15}$
\vspace{5mm}

{$\hat M_{\mathop{\rm{max}}} = a + b \log P + c (B-V)_{\mathop{\rm{max}}}
+ d (M_{\mathop{\rm{mean}}} - M_{\mathop{\rm{max}}})$}
\vspace{2mm}

\begin{tabular}[t]{crrcrr}
& \hspace{6mm} ${\hat{\beta}}$ \hspace{2mm}
& \hspace{7mm} ${\sigma}_{\hat{\beta}}$ \hspace{1mm}
& \hspace{6mm} {\footnotesize{{$\log P(\beta = 0)$}}} \hspace{2mm}
& \hspace{7mm} ${\sigma}_{{\hat M}_*}$ \hspace{1mm}
& \hspace{4mm} $\Delta$ \hspace{4mm}\\ \hline
a & -2.43 & 0.18 & & 0.224 & 10.33\\
b & -3.12 & 0.21 & $< -10$ & & \\
c &  2.23 & 0.37 & -6.22 & & \\
d &  0.53 & 0.40 & -1.01 & & \\ \hline
\end{tabular}
\vspace{0.8mm}

partial SMCC =  0.232 \hspace{1.5cm}
P($H_0$) = $0.193$
\vspace{8mm}

\clearpage
\newpage

\begin{center}
{\bf{TABLE 9 continued}}
\end{center}
\vspace{1mm}

\noindent
{\footnotesize{\underline{39 stars}}}
\vspace{2mm}

{$\hat M_{\mathop{\rm{max}}} = a + b \log P + c (B-V)
+ d (M_{\mathop{\rm{mean}}} - < M_{\mathop{\rm{mean}}}>)$}
\vspace{2mm}

\begin{tabular}[t]{crrcrr}
& \hspace{6mm} ${\hat{\beta}}$ \hspace{2mm}
& \hspace{7mm} ${\sigma}_{\hat{\beta}}$ \hspace{1mm}
& \hspace{6mm} {\footnotesize{{$\log P(\beta = 0)$}}} \hspace{2mm}
& \hspace{7mm} ${\sigma}_{{\hat M}_*}$ \hspace{1mm}
& \hspace{4mm} \hspace{4mm}\\ \hline
a & -4.22 & 0.85 & & 0.090 & \\
b & -1.07 & 0.14 & -8.26 & & \\
c &  1.02 & 0.12 & -9.47 & & \\
d &  0.72 & 0.05 & $<$ -10 & & \\ \hline
\end{tabular}
\vspace{0.8mm}

partial SMCC = 0.925 \hspace{1.5cm}
P($H_0$) = $3.0 \times 10^{-16}$
\vspace{5mm}

{$\hat M_{\mathop{\rm{max}}} = a + b \log P + c (B-V)_{\mathop{\rm{max}}}
+ d (M_{\mathop{\rm{mean}}} - M_{\mathop{\rm{max}}})$}
\vspace{2mm}

\begin{tabular}[t]{crrcrr}
& \hspace{6mm} ${\hat{\beta}}$ \hspace{2mm}
& \hspace{7mm} ${\sigma}_{\hat{\beta}}$ \hspace{1mm}
& \hspace{6mm} {\footnotesize{{$\log P(\beta = 0)$}}} \hspace{2mm}
& \hspace{7mm} ${\sigma}_{{\hat M}_*}$ \hspace{1mm}
& \hspace{4mm} $\Delta$ \hspace{4mm}\\ \hline
a & -2.40 & 0.18 & & 0.22 & 10.2\\
b & -3.19 & 0.17 & $<$ -10 & & \\
c & 2.16 & 0.36 & -6.33 & & \\
d & 0.68 & 0.30 & -1.79 & & \\ \hline
\end{tabular}
\vspace{0.8mm}

partial SMCC =  0.352 \hspace{1.5cm}
P($H_0$) = $3.2 \times 10^{-2}$
\end{document}